\documentclass[lettersize,journal]{IEEEtran}
\usepackage{amsmath,amsfonts}
\usepackage{algorithmic}
\usepackage{algorithm}
\usepackage{array}
\usepackage[caption=false,font=normalsize,labelfont=sf,textfont=sf]{subfig}
\usepackage{textcomp}
\usepackage{stfloats}
\usepackage{url}
\usepackage{verbatim}
\usepackage{float} 
\usepackage{amsmath}
\usepackage{graphicx, subfig}
\usepackage{tabularx}
\usepackage{booktabs}
\usepackage{makecell}
\usepackage{cite}
\usepackage{xcolor}
\usepackage{booktabs}
\usepackage{multirow}
\begin{document}

\title{ Enhancing Throughput for TTEthernet via Co-optimizing Routing and Scheduling:  An Online Time-Varying Graph-based Method}
\author{Yaoxu He, Hongyan Li,~\IEEEmembership{Member,~IEEE,}, Peng Wang
\thanks{}
\thanks{}}

\markboth{Journal of \LaTeX\ Class Files,~Vol.~14, No.~8, August~2021}%
{Shell \MakeLowercase{\textit{et al.}}: A Sample Article Using IEEEtran.cls for IEEE Journals}


\maketitle

\begin{abstract}

Time-Triggered Ethernet (TTEthernet) has been widely applied in many scenarios such as industrial internet, automotive electronics, and aerospace, where offline routing and scheduling for TTEthernet has been largely investigated. However, predetermined routes and schedules cannot meet the demands in some agile scenarios, such as smart factories, autonomous driving, and satellite network switching, where the transmission requests join in and leave the network frequently. Thus, we study the online joint routing and scheduling problem for TTEthernet. However, balancing efficient and effective routing and scheduling in an online environment can be quite challenging.



To ensure high-quality and fast routing and scheduling, we first design a time-slot expanded graph (TSEG) to model the available resources of TTEthernet over time. The fine-grained representation of TSEG allows us to select a time slot via selecting an edge, thus transforming the scheduling problem into a simple routing problem. Next, we design a dynamic weighting method for each edge in TSEG and further propose an algorithm to co-optimize the routing and scheduling. Our scheme enhances the TTEthernet throughput by co-optimizing the routing and scheduling to eliminate potential conflicts among flow requests, as compared to existing methods.   
The extensive simulation results show that our scheme runs $>400$ times faster than standard solutions (i.e., ILP solver), while the gap is only 2\% to the optimally scheduled number of flow requests. Besides, as compared to existing schemes, our method can improve the successfully scheduled number of flows by more than 18\%.

\end{abstract}

\begin{IEEEkeywords}
Time-Triggered Ethernet, online joint routing and scheduling, time-slot expanded graph, dynamic weighting.
\end{IEEEkeywords}

\section{Introduction}

\IEEEPARstart{T}{ime-Triggerd Ethernet}(TTEthernet) provides deterministic latency, deterministic latency jitter, and high-reliability transmission for industrial Internet of Things\cite{ref40}, Internet of Vehicles\cite{ref41}, Augmented Reality (AR)\cite{ref42}, and Virtual Reality (VR)\cite{tnsm2021}, where traffic routing and scheduling serves as a building block, as specified in the standard SAE AS6802\cite{ref1}. Many efforts \cite{ref10,ref11,ref12,ref13,ref16,ref17,ref18,ref19,ref2,ref21,ref22,ref24} study the offline routing and scheduling problem for TTEthernet, since the periodic flow transmission always last for a long time in such TTEthernet scenarios. However, an increasing number of new TTEthernet applications \cite{ref44,ref45,ref46,ref47,ref38,G2019,P2018,T2011} have appeared yielding for online routing and scheduling, where the data transmission requests frequently join in and leave the network. For instance, the development of 5G technology gives rise to new applications of the industrial Internet, including visual inspection with AI\cite{ref44}, fault diagnosis of machines\cite{ref45}, intelligent factory logistics\cite{ref46} and unmanned intelligent inspection\cite{ref47}. Let us take unmanned intelligent inspection as an example, multiple sensors deployed on static devices and mobile robots work together to provide global information for inspection. During the process, the sensor deployed on the robot takes a picture wherever the robot goes and transmits the picture to certain processing nodes immediately\cite{ref38}, otherwise, production accidents can be caused due to outdated picture information. In such a scenario, the picture transmission request frequently joins and leaves, where offline routing and scheduling is incapable of arranging such agile flow requests. On the other hand, some TTEthernet applications inherently require flow requests to be deployed within seconds \cite{G2019} or even milliseconds \cite{P2018,T2011}, yielding routing and scheduling algorithms that can run fast. As a result, online routing and scheduling are becoming increasingly important for TTEthernet.  

Balancing the effectiveness and efficiency of online routing and scheduling for TTEthernet can be quite challenging. Ideally, during a given time horizon, online routing and scheduling can complete the same number of transmission requests which the offline routing and scheduling can at most finish. One straightforward approach is to release all resources occupied by existing transmission requests and reschedule them offline whenever new transmission requests arrive. However, the offline routing and scheduling is NP-hard as in \cite{ref2}. Thus, efficiency cannot be ensured in this manner. Alternatively, to ensure efficiency, one can greedily assign currently available resources to the newly arrived transmission request to meet its QoS requirements without rescheduling previous transmission requests. However, future transmission requests may be blocked by the current greedy resource assignment.  


A promising way to address the efficiency problem (i.e., solving the target problem fast) is to use efficient time-varying graph technology. Importantly, time-varying graphs essentially have the strength to represent resource status in each time slot for a network\cite{wpwcl,wptwc,infocom,shikeyitvt,zttwc}, while the scheduling and routing problem in TTEthernet involves resource assignment over different time slots. Thus, by representing with a time-varying graph, routing and scheduling in TTEthernet (i.e., choosing which link and deciding when to transmit data), can be simply merged into a routing problem (i.e., choosing which edges to transmit data), since the edges in a time-varying graph carry the timing information regardless of their weights. Moreover, existing efforts \cite{ref13,ref16,ref17,ref18,ref2} model the routing and scheduling problem in TTEthernet as an integer linear programming~(ILP) one. A standard way to solve this problem is to use an ILP solver, which lacks scalability as reported in \cite{ref13,ref16,ref17,ref18,ref2}.  We observe, however, that employing a graph-based algorithm to solve the linear programming problem or integer linear programming is consistently much faster than using a solver as reported in \cite{wptwc, infocom, wangpengglobecom, shikeyitvt, guobinquan, zhoudi, jiaziye}. This is because a graph represents the network in a sparse manner. Hence, to identify a path in a graph, a graph-based algorithm only needs to consider the neighbor nodes for each node, while mathematical programming may need to traverse all the node pairs, even including those without connected edges in the network. As a result, solving the online routing and scheduling problem in TTEthernet can benefit from the temporally fine-grained while spatially sparse representation of the time-varying graph.  

\begin{figure*}[htbp]
	\centering
	\includegraphics[width=\linewidth]{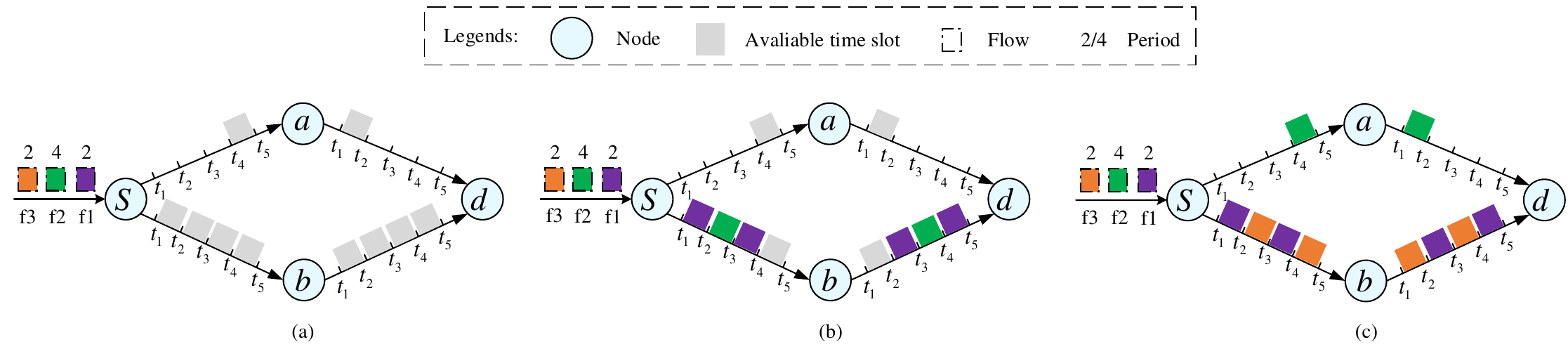} 
	\caption{An example of the effect of this coupling relationship on the completed number of transmission requests.}
	\label{insight}
\end{figure*}

Besides efficiency, we identify that the capability of co-optimization of routing and scheduling becomes a key design factor that can significantly affect the routing and scheduling quality (i.e., the successfully scheduled transmission requests) in TTEthernet. That is, the co-optimization of routing and scheduling can further enhance the schedulability of the problem to eliminate potential conflicts among certain flow requests, as compared to separate optimization of routing and scheduling as reported in \cite{ref6,ref7}. We use Fig. \ref{insight} to illustrate the effect of this capability on the routing and scheduling quality.  Fig. \ref{insight}(a) shows the available transmission resources of a 4-node network within 4 time slots. We assume the network already transmits some background TTEthernet data flow in the network, resulting in the unavailability of links during some time slots. The delay of each link is 1 time slot and consists of the transmission delay. Specifically, the circles and directed links between circles represent nodes and corresponding communication links, respectively. We label 5 time moments on each edge to represent 4 time slots. Each grey block residing within a time slot means the communication link can transmit data within that time slot.  Each grey block appears every 4 time slots.   
The scheduler needs to process three flow requests, represented by purple (f1), green (f2), and yellow blocks (f3), respectively. Flow requests f1, f2, and f3 share a common destination $d$ and will successively arrive at node $s$ every 2, 4, and 2 time slots, respectively. Their maximum allowed end-to-end delays are 4, 8, and 4 time slots. Note that the scheduler cannot know the flow information (e.g., period) in advance until it arrives at $s$. 

Given such a network and flow requests, as shown in Fig. \ref{insight}(b), existing schemes \cite{ref6,ref7} separate the routing and scheduling. They select a path $(s,b,d)$ first. When f1 arrives, they assign the time slots $[t_0,t_1]$ and $[t_2,t_3]$ of link $(s,b)$ and time slots $[t_1,t_2]$ and $[t_3,t_4]$ of link $(b,d)$ to transmit f1.
Hereafter, we can identify a potential space conflict along this path, which is assigning a flow with 4 time-slots period over path $(s,b,d)$ can block the deployment of a potential flow with a period of 2 time-slots. 
The scheduler reported in \cite{ref6,ref7} ignores such a potential conflict, it just assign $[t_1,t_2]$ of $(s,b)$ and $[t_2,t_3]$ of $(b,d)$ to transmit $f_2$. Thus, when f3 arrives, the scheduler in \cite{ref6,ref7} finds that the network cannot accommodate f3 and will reject it. 
Actually, we can transmit all three requests by deploying f2 over path $(s,a,d)$, as shown in Fig. \ref{insight}(c), such that a potential flow with a 2 time-slots period will not be blocked and can be deployed over path $(s,b,d)$.  As a result, co-optimizing the routing and scheduling is promising to enhance the network throughput by eliminating the potential conflicts between flow requests.  

There are already efforts\cite{ref10,ref11,ref12,ref13,ref16,ref17,ref18,ref19,ref2,ref21,ref22,ref24,ref6,ref7,ref23,ref29} studying routing and scheduling problem in TTEthernet. Among them, many efforts \cite{ref10,ref11,ref12,ref13,ref16,ref17,ref18,ref19,ref2,ref21,ref22,ref24} study the offline routing and scheduling in TTEthernet which essentially run slowly and cannot operate in online environments. On the other hand, some efforts \cite{ref6,ref23,ref29} study the separate routing and scheduling optimization in online environments without considering potential conflicts between flow requests, suffering from poor schedule quality.
Specifically, the works reported in \cite{ref23,ref29} just greedily find one optimal (e.g., minimum hops, minimum delay) path for current flow requests, which can easily lead to premature congestion in the path, resulting in a sharp decline in schedule quality. Instead of searching for the optimal path, reference \cite{ref6} selects the path according to the number of hops and the number of occupied time slots on the link to achieve load balancing. In addition, all three efforts can block resource assignments for potential flow requests.
To overcome this bottleneck, reference \cite{ref7} identifies the potential conflict relationship among flow requests, however, it just predetermines the path to transmit the flow before scheduling. The separate routing and scheduling reduces the capability of a network to eliminate the conflicts among flows as shown in Fig. \ref{insight}, causing poor schedule quality. As far as we know, no previous works focus on co-optimizing the routing and scheduling to increase the chances of eliminating the conflicts among flows.   

To achieve high-quality and fast routing and scheduling in TTEthernet,
we first design a time-varying graph to model the network.
Next, we model the problem as an ILP based on the time-varying graph, providing an upper bound to evaluate our solution. Then, we design a dynamic weighting method to reveal potential conflicts among flow requests. The graph and weighting method allows us to co-optimize routing and scheduling by simply designing a routing algorithm. Finally, we conduct experiments to prove the performance of our scheme. The specific contributions of the paper are as follows:
\begin{enumerate}
	\item We design a time-slot expanded graph (TSEG), which can accurately represent the resources of TTEthernet in both spatial and temporal dimensions.
    In particular, we design a type of edge called the inter-hyper-period edge to support the correct scheduling of flows arriving at different times.
    Each edge in TSEG essentially carries timing information. Thus, choosing an edge in TSEG is equivalent to choosing which link and when to transmit data, enabling us to solve the routing and scheduling problem in TTEthernet as a pure edge selection problem. 
	\item To rigorously study the potential gain of our scheme, we model the routing and scheduling problem in TTEthernet as an ILP based on TSEG, maximizing the successfully scheduled flow requests via optimizing the decision variable of choosing which link and when to transmit data. As compared to existing problem formulation \cite{ref16}, our formulation significantly reduces the problem size by merging the route variables and time slot selection variables into a single variable. This model provides rigorous constraints to help design our scheme. Particularly, it can be used to evaluate the optimality gap and justify why we use TSEG as a standard baseline method. 
	\item We design a dynamic weighting method for each edge in TSEG to reveal potential conflict relationships among different flow requests. In particular, it can be observed that selecting the path with the minimum total weight among all paths can minimize the impact of current flow on future flow. Based on this observation, we further propose a joint routing and scheduling algorithm for each arrival flow, which finds the path with minimum weights while satisfying the requirements of flows.
	\item We conduct experiments to evaluate our proposed scheme using the topology of real-world industrial networks. Evaluation results show that our scheme can significantly improve the runtime and schedule quality as compared to existing schemes. Specifically, our approach runs $>400$ times faster than a standard ILP solver, while the gap of the successfully scheduled number of flows between our scheme and the optimal ILP solver is only 2$\%$. Besides, our scheme can schedule 18\% more flows than baselines on average. 
\end{enumerate}
This paper is structured as follows. The related work is presented in Section \ref{Related Work}. Section \ref{System Model} discusses the network model, the definition of basic concepts in TTEthernet, and the network configuration and management model. Then, the time-slot expanded graph (TSEG) is designed. Based on TSEG, Section \ref{model} introduces ILP formulations for routing and scheduling problem with the optimization objective of maximizing the number of scheduled flows. The online joint routing and scheduling scheme is designed to minimize the impact of resource allocation for current flow on future flow in Section \ref{Time-Varying Graph-Based Algorithm}. The presented scheduling approaches are evaluated in Section \ref{SIMULATION RESULTS}. Finally, we conclude this paper in Section \ref{Conclusion}.

\section{Related Work}\label{Related Work}
The routing and scheduling problems in time-triggered networks have been extensively studied. 
We roughly divide the current efforts studying this problem into four categories \textemdash online and offline routing and scheduling, co-optimization, and separate optimization of routing and scheduling.  

We first introduce the differences among these categories. Online routing and scheduling refers to assigning resources to a  flow after it joins a network. In an online environment, the flows frequently join, invoking the scheduler every time a new flow joins. On the other hand, the offline routing and scheduling assign resources for a bunch of flows in advance. There tends to be no new flow after the system begins to work. Generally, \textcolor{black}{offline} routing and scheduling take much more time than online routing and scheduling. 
Regarding optimization of routing and scheduling, separate optimization finds one path first and then decides when to transmit data along each link of that path. This hierarchy method can split the original problem into a routing problem and a scheduling problem. In this manner, improper path choices can be made without carefully considering the available resources, 
leading to sub-optimality.  Co-optimizing the routing and scheduling, however,   maintains the optimality, by considering when to transmit data while deciding over which link to transmit data.           

\subsection{Online Routing and Scheduling}

\subsubsection{Separate optimization}



Many efforts \cite{ref23,ref7,ref6,ref29,ref30,ref31,ref32} study the separate routing and scheduling optimization in online environments, leading to sub-optimal results. For instance, the authors in \cite{ref6} propose a standard separate optimization method. They first identify a set of paths, evaluating the load status of each path using the number of occupied time slots of each link and the number of hops. Next, they always choose the path with the lightest load and answer when to transmit data along the path by solving an Integer Linear Programming problem. Obviously, a path able to transmit a small period flow has more resources than a path that can only transmit large period flows, thus carrying less load. With this observation, reference \cite{ref6} will arrange a flow with a large period over a path that can transmit a small period flow. Since TTEthernet has more chances of transmitting large period flows than small period flows, their load balancing way will in great chances weaken the ability of TTEthernet to transmit the small period flows, leading to sub-optimal results.

Besides, to propose a fast online routing and scheduling method, reference \cite{ref23} finds the shortest path for each arrived flow while reference \cite{ref29} models the scheduling problem as a vector packing problem and designs four heuristic algorithms to find the ideal path. 
These methods do propose fast ways to find a solution, with the schedule quality being compromised by separate routing and scheduling. Moreover, the work in \cite{ref7} uses deep neural networks (DNN) to solve the online routing and scheduling problem, adapting to large-scale networks. However, this method requires a large number of samples to train the DNN with the schedule quality highly dependent on the training samples. Moreover, references \cite{ref30,ref31,ref32} consider the online routing and scheduling involved with network topology changes. This topic falls out of the scope of our work, as we consider a wired TTEthernet, where we assume no topology change. 


\subsubsection{Co-optimization} 
 Efforts \cite{ref5,ref28} study the co-optimization of routing and scheduling in an online environment. The Work in \cite{ref5} models online joint routing and scheduling in time-sensitive networks as integer linear programming problems. These models cannot be applied to TTEthernet since they adopt a coarse-grained timing concept which allows an MTU-sized frame to traverse the whole network within a time slot, while in TTEthernet, one frame can at most traverse one link within one single time slot as in \cite{ref10,ref16,ref18,ref6,ref7}.  Besides, the authors in  \cite{ref28} solve the online joint routing and scheduling problem for multicast by extending the original integer linear programming into cluster linear programming, to accelerate the scheduling speed. However, the authors reschedule previous flows with the newly arrived flow, which can be quite time-consuming when the number of flows is huge. Besides, the change of resource assignment for previous flows will incur high configuration overhead.

\subsection{Offline Routing and Scheduling}
\subsubsection{Separate optimization} Offline separate routing and scheduling refers to computing the paths for multiple flows first and allocating available resources of each path to multiple flows, which can be reduced to the well-known NP-complete bin-packing problem as in \cite{ref10}. Despite its NP-completeness in the scheduling phase, optimally solving this separate routing and scheduling problem does not necessarily lead to a good schedule quality, since the chosen path for a flow may not be the optimal path in the first place. There are efforts to solve this problem using satisfiability modulo theories (SMT) based scheme\cite{ref11}, mixed integer linear programming (MILP) based scheme \cite{ref12} and integer linear programming based scheme \cite{ref13}, suffering from long running time.  

\subsubsection{Co-optimization of offline routing and scheduling}
Many existing works study the co-optimization of routing and scheduling \cite{ref16,ref17,ref18,ref19,ref2,ref21,ref22,ref24}, which is proved to be an NP-hard problem \cite{ref2}. In fact, solving such a problem could provide us an upper bound on at most how many flows can be accommodated in TTEthernet, however, the solution suffers from long running time and cannot be applied to online routing and scheduling.  

Some efforts \cite{ref2,ref16,ref17,ref18} choose to model the problem as integer linear programming and use a solver to solve it, which could suffer from long running time when the network scale and the number of flows is large. For instance, Schweissguth et al. \cite{ref16} first formally model the joint routing and scheduling problem as an ILP model. Later, the authors in \cite{ref17} extend the model to solve the problem in a multi-cast scenario. To improve the running speed, reference \cite{ref18} reduces the problem size by separating the flows into different groups, optimally assigning resources for each group of flow via repeatedly solving the integer linear programming, still suffering from long running time when the network scale is large.

Other efforts resort to using heuristics to speed up solving the problem\cite{ref21,ref22,ref19,ref24}, compromising the schedule quality. Reference \cite{ref21} have proposed a Tabu Search-based approach to calculate the schedule table of time-critical flows while minimizing the end-to-end latency of RC traffic, while the authors in \cite{ref22} proposed a genetic algorithm to solve the scheduling problem with built-in compression. Besides, the authors in \cite{ref19} solve the offline problem in an online manner, i.e., choosing to assign resources for the given flows one by one in the order of their assigned scheduling priority weight.  
Besides, the authors in \cite{ref24} identify all possible paths for each source and destination set in advance and then determine the time slot that satisfies the shortest transmission delay for each flow. In fact, the time complexity for enumerating all the paths can be exponential as in \cite{allpath2001}. 

\section{System Model}\label{System Model}
\subsection{Basic definitions}
\subsubsection{Network model}
The notations used in this paper are listed in Table \ref{tab:example}. Without loss of generality, we consider a time-triggered Ethernet $G=\{\mathcal{V},\mathcal{E}\}$, where $\mathcal{V}$ represents network nodes, consisting of switches and end systems, while $\mathcal{E}$ collects all the communication links connecting nodes. Note that, each communication link 
is full-duplex. We denote each communication link in TTEthernet via two directional communication links $(u,v)$ and $(v,u)$, respectively. Obviously, directional link $(u,v)$ represents that $u$ can transmit data to $v$.
In such a network, we ask along which link and when to transmit each arrived flow $f$, such that the network can accommodate as many flows as possible with each flow's requirements met.  

\subsubsection{Definitions of Time slot and Flow }
Let us first consider transmitting one flow request $f$ in a TTEthernet $G$ within a time horizon $T$.
Following works \cite{ref10,ref16,ref18,ref6,ref7}, we divide a given time horizon $T$ into multiple equal-length time periods\textemdash  time slot $\tau$, simplifying the routing and scheduling problem in TTEthernet from a continuous-time problem into a discrete-time one. Since the maximum tolerable delay and transmission frequency (i.e., period) in TTEthernet is constrained on each frame of a flow request $f$, we let the length of a time slot\textemdash $|\tau|$ be the maximum delay experienced by an MTU-sized frame (i.e., a frame with Maximum Transmission Unit\textemdash \textcolor{black}{1500} bytes)\cite{ref13} to cross a link in TTEthernet following work \cite{ref10,ref16,ref18,ref6,ref7}. We claim that this setting is reasonable since TTEthernet is typically deployed in a relatively small geographic area\cite{ref40,ref41,ref42,tnsm2021,ref1}, where the propagation delay and transmission rate of each link differs little.  As a result, any frame can traverse only one communication link within each time slot.


We formally define a flow $f$ in TTEthernet as $f=\{t_f,s_f,d_f,p_f,m_f\}$.  Specifically, as in \cite{ref10},  after the flow $f$ arrives in the network at $t_f$, TTEthernet starts to transmit the flow in frames from $f$'s source node $s_f\in\mathcal{V}$ to destination node $d_f\in\mathcal{V}$ repeatedly in a period a $p_f$ time slots. Within each period, TTEthernet only transmits one frame of the flow.  In addition, TTEthernet constrains each frame to be transmitted within a maximum tolerable time duration (i.e., delay) which can at most reach $m_f$ time slots. The duration starts when $s_f$ starts to send the frame and ends when $d_f$ receives the frame.

With the above definitions,  our considered problem for transmitting $f$ asks on which link and within which time slot to transmit data, such that the transmission frequency ($p_f$) and maximum tolerable delay ($m_f$) of $f$ is met. In addition,  it is not practical to compute the routing and scheduling for each period of $f$, we assume the resource assignment results within each period for a flow are exactly the same as in \cite{ref10}. 

\subsubsection{Definition of Hyper-period}
Now we are ready to discuss the time horizon $T$ within which resource scheduling should take place. A straightforward way is to schedule each flow within a single period of $f$ (i.e., within $p_f$ time slots), and then update the available resources of TTEthernet in each $p_f$ time slots. 
Recall that we will assign resources for multiple flows, which could have different transmission periods. Hence, updating the available network resources for flows with different periods will be highly complicated.  

To address this issue, a common way is to introduce a finite time horizon\textemdash \textit{hyper-period} as our considered time horizon $T$  \cite{ref10,ref16,ref18,ref6,ref7}. Specifically, we set the length of a hyper-period as $N$ time slots, which is the least common multiple of all the period values of each potential flow \footnote{People can refer to the historical flow information in a TTEthernet as the period information of potential flows} in TTEthernet. 
To be more specific, let $T$ contains $N$ time slots with $T=\{\tau_1,...\tau_N\}$,  within which the data of any potential flow of TTEthernet will be transmitted at least once, since the length of $T$ is integral times of each potential flow's period. As a result, as long as we have determined the routing and scheduling for a flow $f$ in $T$, the network can reuse the result every $N$ time slots and TTEthernet's available network resources will be the same every time duration $T$ (i.e., $N$ time slots).  Therefore, maintaining the resource status will be much easier using hyper-period $T$. 

\subsubsection{Timing Relationship}
\begin{figure}[htbp]
	\centering
	\includegraphics[width=0.5\textwidth]{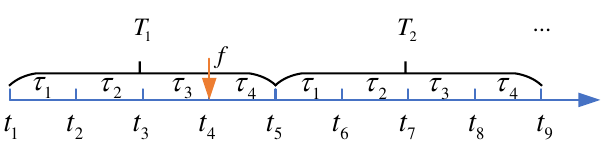} 
	\caption{An example of the hyper-period in an online scenario} 
	\label{hyper}
\end{figure}

With the concept of hyper-period, we only need to maintain the resource status of TTEthernet in $T$ and schedule flow within $T$. However, when a flow arrives at $t_f$, we still cannot know where $t_f$ is located in such a $T$. To overcome this, we define a time moment when the TTEthernet starts to work as \textcolor{black}{$t_1$}. From there, we can encode the time dimension with time slot $\tau$ (e.g., $\tau_i$ represents $i_{th}$ time slot from \textcolor{black}{$t_1$}). We merge $N$ time slots as one hyper-period $T$. As a result, we can also encode the time dimension with $T$ (e.g., $T_i$ represents the $i_{th}$ time hyper-period for \textcolor{black}{$t_1$} ) as shown in Fig. \ref{hyper}. With such a definition, we can use the network resource status depicted in $T_i$ to schedule $f$ iff $t_f\in T_i$. Note that, maintaining the network status within a certain hyper-period is easy, since the network resource status of any two adjacent hyper-periods (e.g., $T_i$ and $T_{i+1}$) is the same if no new flow joins or old flow leaves.

Particularly, the flow $f$ can arrive at the middle of a time horizon $T$, say the $i_{th}$ hyper-period $T_i$, we still can use the information of network resource in $T_i$
to schedule $f$, since the resource availability within a hyper-period $T$ appears periodically with a period of $T$. For instance, as shown in Figure \ref{hyper}, suppose a new flow request $f$ arrives at time \textcolor{black}{$t_4$} of $T_1$ with a period of 4 time slots. To schedule such a flow,  the scheduler needs to take as input the network resource status from \textcolor{black}{$t_4$ to $t_8$}. This means that the transmission of $f$ could potentially cross two adjacent hyper-periods. Actually, $T_1$ and $T_2$ have the same resource status, the resource status \textcolor{black}{from $t_5$ to $t_8$} are actually the same as that from \textcolor{black}{$t_1$ to $t_4$}. Specifically, \textcolor{black}{$t_5=t_1+N\cdot |\tau|$, $t_9=t_5+N\cdot |\tau|$}, where $N\cdot |\tau|$ is the length of hyper-period $T$. In addition, it is obvious that the maintained resource status within $T$ is sufficient to schedule flow $f$ which arrives at the beginning of $T$. As a result, maintaining resource status within $T$ is sufficient to schedule any incoming flow.



\subsubsection{Network Configuration and Management Model}
\begin{figure}[htbp]
	\centering
	\includegraphics[width=0.45\textwidth]{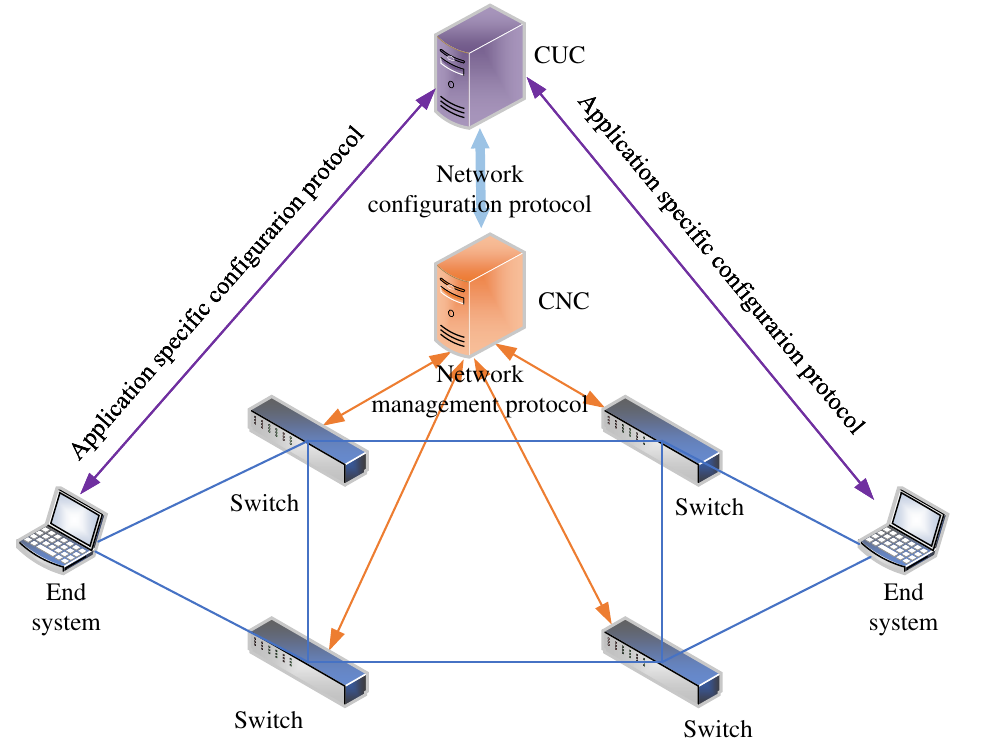} 
	\caption{Centralized Conﬁguration and Management Model} 
	\label{architec}
\end{figure}

We use a centralized network architecture\cite{archit1,archit2,archit3} where the entire network is perfectly synchronized as in \cite{ref6,ref7}, as shown in Fig. \ref{architec}. This architecture consists of two kinds of controller nodes\textemdash Centralized User Conﬁguration (CUC) node and Centralized Network Conﬁguration (CNC) node, end systems, and switches. End systems are data sources generating flow transmission requests with deterministic delay requirements and also destinations receiving data from other end systems. 
In particular, CUCs communicate with end systems to collect their time-triggered flow transmission requests through application-specific configuration protocol
and forward the request to CNC nodes through network configuration protocol. Hereafter, CNCs run the online routing and scheduling algorithm to decide along which link and within which time slot to transmit the frames of the flow request. Next, CNC configures the corresponding switches according to their calculated results through the configuration protocol (e.g., the OpenFlow protocol\cite{openflow2014}. Finally, each switch updates its scheduling table to reserve corresponding resources.
\subsection{Design of Time-Slot Expanded Graph}
Recall that we only need to maintain the resources status of TTEthernet within a hyper-period $T$ to schedule flows. Actually, flows with different periods could require resources from different time slots, resulting in a time-varying TTEthernet within $T$. A straightforward way to represent the time-varying resource status is using a time-varying graph (e.g., time-expanded graph\cite{teg2015}, time-aggregated graph\cite{tag08}) to represent the available resources for each time slot in $T$. However, in this manner, the period (i.e., N time slots) of resource availability in TTEthernet cannot be depicted, causing it difficult to schedule a flow to cross two adjacent hyper-periods. As a result, representing resource status within $T$ via traditional time-varying graphs is insufficient to schedule any flows.

To overcome this issue, we design a time-slot expanded graph $G_T= (\mathcal{V}^{'},\mathcal{E}^{'})$ to represent the available transmission resources within $T$, caching resources within $T$ and the period of $T$ for a TTEthernet network.  
Specifically, 
\subsubsection{Common representation}
Similar to time-expanded graph\cite{teg2015}, for each entity node (i.e., switches and end systems) in TTEthernet,  we generate one vertex in each time slot of $T$. For instance,  we will create $N$ vertices for node $u$, among which the representation of $u$ in a general time slot $\tau_i$ is denoted as $u_i$. We collect all the representation vertices into a vertices set, denoted as $\mathcal{V}'$ to differ from the set $\mathcal{V}$ of entity nodes in TTEthernet.  
Next, for each communication link $(u,v)$ in TTEthernet, if it has not been allocated to any flow for transmission in a general time slot $\tau_i$, we will use a directional edge to connect $u_i$ and $v_i$, representing link $(u,v)$ is available for data transmission from $u$ to $v$ in $\tau_i$. We collect all the transmission edges into edge set  $\mathcal{E}{'}$. Besides the transmission links, for each vertex $u_i$ with $i<N$, we will connect the two temporally adjacent vertices $u_i$ and $u_{i+1}$ with a directional edge $(u_i, u_{i+1})$, \textcolor{black}{representing that data can be cached on node $u$ from $\tau_i$ to $\tau_{i+1}$.} 
We also collect such caching links into edge set $\mathcal{E}'$. 
\subsubsection{Unique representation}
To represent the period of network resource status, we introduce the \textcolor{black}{inter-hyper-period edges}, whose direction is from the last time slot of the represented hyper-period to the first time slot of the hyper-period. Specifically,  for each vertex $u_N$ within $\tau_N$, we connect $u_{N}$ to $u_1$ with a directional edge $(u_N,u_1)$. With such \textcolor{black}{inter-hyper-period edges},  we can schedule flow requests arriving within $T$ to cross at most $N$ time slots, via crossing the  \textcolor{black}{inter-hyper-period edges}. Note that, such a scheduling is feasible since the resource availability has periods of $N$ time slots. We collect such  \textcolor{black}{inter-hyper-period edges} into edge set $\mathcal{E}'$. With the help of such edges, maintaining the resources within $T$ can be sufficient to complete the scheduling of flow requests arriving at any time slot.

\underline{\textit{Remark:}} In fact, our proposed time-slot expanded graph is very different from the time-expanded graph, despite that they have similar structures. Recall that we define the length of the time slot to allow an MTU-sized frame to cross any link in TTEthernet. That is, within a time slot, only one frame is allowed to transmit at most one link, different from a traditional time-varying graph where data can traverse multiple edges within one snapshot (i.e., one time slot). Again, due to the small length of a time slot, the number of frames one node needs to cache from one time slot to a future time slot is also very small (depending on the number of incoming links of that node), hence, we assume the storage capacity of each node is infinite in time-slot expanded graph (as compared to the number of frames they need to store), while the storage capacity of nodes is always limited in traditional time-expanded graph\cite{teg2015}. In particular, our considered time-slot expanded graph supports flow to traverse two adjacent hyper-periods via maintaining the resource status of one hyper-period, allowing us to solve the routing and scheduling problem in TTEthernet as a pure edge selection problem.

\begin{figure}[htbp]
	\centering
	\includegraphics[width=0.4\textwidth]{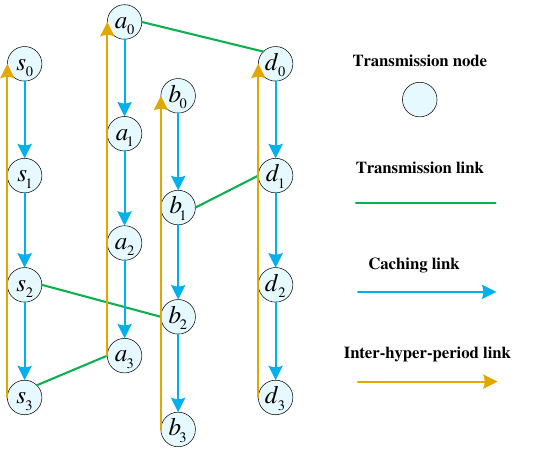} 
	\caption{Time-slot expanded graph model} 
	\label{Fig1}
\end{figure}


We give an example to further illustrate how to use TSEG to schedule a flow. 
We first show the TSEG within $T=\{\tau_1,\tau_2,\tau_3,\tau_4\}$ in Fig. \ref{Fig1}. 
Suppose a flow request $f=\{t_1,s_f=s,d_f=d,p_f=4, m_f=4\}$ arrives at \textcolor{black}{$t_1$} (i.e., within $\tau_1$) on node $s$ with a destination $d$. Its period and maximum tolerable delay are both four time slots. 
As shown in TSEG, we can choose either the edge set (i.e., path) $\{(s_3,a_3),(a_3,a_0),(a_0,d_0)\}$ or $\{(s_2,b_2),(b_2,b_3),(b_3,b_0),(b_0,b_1), (b_1,d_1)\}$ to transmit the flow $f$, while satisfying its period and delay requirements. Specifically, we take path $\{\{(s_2,b_2),(b_2,b_3),(b_3,b_0),(b_0,b_1), (b_1,d_1)\}$ as an example. It means scheduler transmits data over link $(s,b)$ in $\tau_3$, and 
stores the data on node $b$ from $\tau_4$ to $\tau_6$ (i.e., presented as $\tau_2$), and then use link $(b,d)$ to transmit the frame within $\tau_7$ (i.e., presented as $\tau_3$). Along the path, edge $(b_2,b_3)$ is no longer a caching but a plain edge to connect the different slots. Edge $(b_2,b_3)$, together with edge $(s_2,b_2)$ means transmitting the frame from $s$ to $b$ requires the whole $\tau_3$, while edge $(b_3,b_0)$ and $(b_0,b_1)$ represent storing the frame from a time slot to its adjacent time slot. In particular, here $(b_3,b_0)$ also means storing the frame from the current hyper-period (i.e., $\tau_4$) to the adjacent hyper-period (i.e., $\tau_5$). As a result, the inter-hyper-period edge $(b_3,b_0)$ connects two sub-paths (i.e.,$\{(s_2,b_2),(b_2,b_3)\}$ and $\{(b_0,b_1), (b_1,d_1)\}$ ) of two adjacent hyper-period into one end-to-end path.

\IEEEpubidadjcol

\begin{table}[!htbp]
	\centering
	\caption{Notations used in this paper.}
	\label{tab:example}
	\resizebox{0.5\textwidth}{!}{
	\begin{tabular}{|c|l|}
		\hline
		\textbf{Notations} & \textbf{Description}\\
		\hline
		$\mathcal{V}$ & the set of nodes for time-triggered Ethernet. \\ 
        \hline
		$\mathcal{E}$ & the set of links for time-triggered Ethernet. \\
        \hline
            $(u,v)$ & The directional communication link from node $u$ to node $v$ in $\mathcal{E}$. \\
        \hline
		$T$ & the finite time horizon \textemdash hyper-period for scheduling problem. \\
        \hline
		$\lvert\tau\rvert$ & the length of a time slot. \\
        \hline
            $t_f$ & the arrival time of flow $f$. \\ 
        \hline
            $s_f$ & the source node of flow $f$. \\ 
        \hline
            $d_f$ & the destination node of flow $f$. \\ 
        \hline
            $p_f$ & the period of flow $f$. \\ 
        \hline
            $m_f$ & the maximum tolerable delay of flow $f$. \\ 
        \hline
            $N$ & The number of time slots contained in $T$. \\ 
        \hline
		$G_T$ & \makecell[l]{the time-slot expanded graph (TSEG) \\ for representing time-varying resource status of network $G$.} \\
        \hline
        $\mathcal{V}^{'}$ & the set of vertices in TSEG. \\
        \hline
	$\mathcal{E}^{'}$ & the set of edges in TSEG. \\
        \hline
        $u_i$ & the vertice representation of node $u$ in time slot $\tau_i$. \\
        \hline
        $(u_i,v_i)$ & the directional edge to connect vertices $u_i$ and $v_i$. \\
        \hline
		$x_{i, j, q}^{(k)}$ & \makecell[l]{this decision variable is 1, if flow $f_k$ \\ crosses the transmission edge $(i_q,j_q)$, Otherwise 0.} \\
        \hline
		$x_{i, i, q}^{(k)}$ & \makecell[l]{this decision variable is 1, if flow $f_k$ crosses the caching edge \\ $(i_{q-1},i_q)$ or inter-hyper-period edge $(i_N,i_1)$, Otherwise 0.} \\
        \hline
		$Z^{(k)}$ & \makecell[l]{this decision variable is 1, if flow $f_k$ is successfully \\ scheduled, otherwise it is 0.} \\
        \hline
		$R$ & the number of variables for problem \textbf{TSEG-MAX}. \\
	\hline
            $K$ & the number of constraints for problem \textbf{TSEG-MAX}. \\
        \hline
        $\mathcal{P}_{i,j,q}$ & \makecell[l]{the transmission edge $(i_q,j_q)$'s potentially supported \\ flows' period set.} \\
	\hline
        $w_{i,j,q}$ & \makecell[l]{the transmission capability for transmission edge  $(i_q,j_q)$ \\ in TSEG.} \\
        \hline
        $\mathcal{E}'_f$ & the transmission edges set traversed by flow $f$. \\ 
        \hline
        $\mathcal{E}''$ & \makecell[l]{the transmission edges set whose weight need \\ to be updated after new flow $f$ is scheduled.} \\ 
        \hline
        $Q$ & a queue collecting all the vertices in TSEG initially. \\ 
        \hline
        $\mathcal{H}(u_q)$ & the smallest weight of the path from source node $s$ to $u_q$. \\ 
        \hline
        $\mathcal{R}(u_q)$ & the penultimate hop node of the path from source node $s$ to $u_q$. \\ 
        \hline
        $\mathcal{O}$ & \makecell[l]{the path set that records the minimum weight path starting \\ from each slot within $p$ for new flow $f$.} \\ 
        \hline
	\end{tabular}
 }
\end{table}

%

\section{Problem formulation} \label{model}
We consider an online environment where the flow dynamically joins and leaves. We aim at jointly routing and scheduling each arrived flow in a proper manner, such that the TTEthernet can accommodate as many flows as possible. 
Within an observation time range $T'$ ($T'>>T$), assuming $T'$ starts from $t_1$, we define $F$ as the set of flows that will arrive within $T'$. Note that, we cannot know the information of incoming flow until it arrives. For a general $k_{th}$ arrival flow $f_k$ in $F$, 
we assign resources for it within a hyper-period $T=\{\tau_1,...,\tau_N\}$. Before proceeding to the constraints, we first introduce the decision variables $x_{i,j,q}^{k}$, which represent whether the $k_{th}$ arrived flow $f_k$ of $F$ will cross an edge or not. In a special case where $i=j$, equation $x_{i,i,q}^{k}=1$ represents the flow crosses caching edge $(i_{q-1},i_q)$ with $q<N$ or inter-hyper-period edge \textcolor{black}{$(i_{N},i_1)$} with $q=N$, otherwise $x_{i,i,q}^{k}=0$. In addition, $x_{i,i,q}^{k}=1$ with $i\neq j$ represents the flow traverse the transmission edge $(i_q,j_q)$, otherwise $x_{i,i,q}^{k}=0$. 
To ensure feasible routing and scheduling, the variables should satisfy the following constraints:

\subsection{Unique Flow Conservation in TSEG}
Let us consider the transmission of a general flow $f_k\in F$ in TTEthernet. Recall we represent the source and destination node in $V$ via vertices set $s_{f_k}=\{s_1(f_k),...s_N(f_k)\}$ and $d_{f_k}=\{d_1(f_k),...,d_N(f_k)\}$ in $\mathcal{V}'$, except these two vertices set, frames entering a relay node $i$ within a time slot (e.g., entering vertex $i_q$) must leave the node in a later time slot (e.g., leaving vertex $i_{q+1}$). This is due to the definition that a time slot can only support a frame to traverse one link.  Hence, we can divide the flow conservation of a relay vertex into two cases: 

a): If the general relay node $i$ receives the frame of $f_k$ via a transmission edge within a general time slot $\tau_q$, the earliest time slot it can transmit the frame is $\tau_{q+1}$. Due to the special structure of TSEG, the received frame must traverse one caching edge $(i_q, i_{q+1})$ (or $(i_N,i_1)$ with q=N) to be sent in a later time slot than $\tau_q$. Thus, regarding any relay node $i \in \mathcal{V}-s_{f_k}\cup d_{f_k}$ transmitting $f_k$, we have, 
\begin{equation}
\label{eq:input}
 \sum_{a\in \mathcal{V},a\neq i} x_{a,i,q}^{(k)}=x_{i,i,q}^{(k)},\forall f_k \in F, \tau_q\in T.
\end{equation}

b): Naturally, if node $i$ receives the frame of $f_k$ via a caching edge, it can freely send the data to its any outgoing edge. For $i \in \mathcal{V}-s_{f_k}\cup d_{f_k}$ transmitting $f_k$, we also have, 
\begin{equation}
\label{eq:output}
x_{i,i,q-1}^{(k)}=\sum_{b\in \mathcal{V}} x_{i,b,q}^{(k)},\forall k \in F, \forall q\in [2,N].
\end{equation}

\subsection{Common constraints}

 One frame is the minimum unit to be transmitted in TTEthernet. Thus, for each frame of $f_k$, it can only be transmitted along a single path. In addition, regarding flow $f_k$, if we successfully schedule it in $T$, TTEthernet should transmit  $\frac{T}{p_{f_k}}$ frames within $T$ of $f_k$ (recall $p_{f_k}$ is the period of $f_k$). Otherwise, no frames of $f_k$ are transmitted.  As a result, we can at most choose $\frac{T}{p_{f_k}}$ edges in TSEG for transmission, for the source node $s_{f_k}$, namely, 
\begin{equation}
\label{eq:onepath}
\sum_{a\in \mathcal{V},a\neq i}\sum_{\tau_q\in T} x_{i,a,q}^{(k)} \leq \frac{N}{p_{f_k}}, i=s_{f_k}, \forall k \in F.
\end{equation}
This constraint itself is insufficient to ensure sing-path transmission, we will use it together with other constraints in the following to achieve that. 

Regarding the capacity constraints of edges, a transmission edge in TSEG can at most transmit one frame, since each link can only transmit one frame within one time slot according to the definition of the time slot. For any transmission edge in TSEG, we have, 
\begin{equation}
\sum_{k \in F} x_{i, j, q}^{(k)} \leq 1 , \forall i,j \in V, i \neq j, \tau_q \in T.
\label{conc}
\end{equation}
Recall that the caching edges in TSEG have infinite capacity, we omit their capacity constraints here. 

\subsection {QoS requirements}
\subsubsection{Periodical Transmission}
Following work \cite{ref10}, we constrain the flow $f_k$ to traverse the same path in each period. That is, if $f_k$ is transmitted over link $(i,j)$ within time slot $\tau_q$, it must traverse this link within time slot $\tau_{q+ \alpha\times p_{f_k}}$ with any $\alpha \in \mathbb{Z}$ and $1\leq q+\alpha\cdot p_{f_k}\leq N$. 
Thus, regarding any link $ i,j \in \mathcal{V}$ and flow $\forall f_k \in F$, we have,
\begin{equation}
\label{eq:7}
 x_{i,j,q}^{(k)}=x_{i,j,q+ \alpha p_{f_k}}^{(k)}, \tau_q\in T, \{\alpha\in \mathbb{Z}|1\leq q+\alpha p_{f_k}\leq N\}.
\end{equation}
Obviously, this constraint ensures the frame to be transmitted in a period of $p_{f_k}$ time slots. Moreover, constraints (\ref{eq:onepath}),  flow conservation constraints (\ref{eq:input}) and (\ref{eq:output}) and periodical transmission constraints (\ref{eq:7}) ensure the single-path transmission.   

\subsubsection{{Maximum Tolerable Delay}}


With the above periodical single-path transmission, let us consider the maximum tolerable delay constraints of flow $f_k$ within $T$. If $f_k$ is successfully scheduled, it will send a frame every $p_{f_k}$ time slot with each frame's experienced delay not exceeding  $m_{f_k}$. Hence, the total maximum allowed delay within $T$ for $f_k$ adds up to $\frac{N\times m_{f_k}}{p_{f_k}}$. Since we cannot know which time slot is the first time slot to transmit the frame within $T$, we can constrain the total experienced delay of each period to not exceeding $\frac{N\times m_{f_k}}{p_{f_k}}$. From the paths of frames as shown in Fig. \ref{Fig1}, we can easily know that the number of time slots, experienced by a frame traversing a path, equals the number of caching edges/inter-hyper-period edges plus 1. For a flow $f_k$, we have, 
\begin{equation}
\label{eq:e2e}
\sum_{i\in \mathcal{V}}\sum_{q\in T} x_{i,i,q}^{(k)} +  \frac{N}{p_{f_k}} \leq \frac{N\times m_{f_k}}{p_{f_k}}, \forall k \in F.
\end{equation}
It is easy to show that this constraint can also ensure that, the transmission of a frame of $f_k$ falls within the maximum tolerable delay, due to the above periodic transmission constraints and single path transmission constraints. 

 \subsubsection{Indicators for Successful Schedule}
 Here we formally define the variables $Z^{(k)}$ to indicate whether flow $f_k$ is successfully scheduled ($Z^{(k)}=1$) or not ($Z^{(k)}=0$). The scheduler decides to transmit   $\sum_{j \in \mathcal{V}} \sum_{q=1}^{N} x_{i, j, q}^{(k)}$ frames within $T$,  each of which will meet $f_k$'s QoS requirements with above constraints satisfied,  while successful schedule of flow $f_k$ requires to transmit  $\frac{N}{p_{f_k}}$ frames. Thus, we can have, 
\begin{equation}
\label{eq:3}
Z^{(k)} \leq \frac{p_{f_k} \sum_{j \in \mathcal{V}} \sum_{q=1}^{N} x_{i, j, q}^{(k)}}{N},\forall k \in F,i=s_k.
\end{equation}
 
\subsection{Problem Formulation and Description}
With the above constraints, our goal is to maximize the number of successfully scheduled flows in $F$ in TSEG. Thus, we have, 
\begin{equation}
\begin{aligned}
\textbf{TSEG-MAX:} \quad &\max\sum_{k\in F} Z^{(k)}\\
&s.t. (1)-(7).
\end{aligned}
\end{equation}
We model the co-optimization of routing and scheduling problem as an ILP problem based on our TSEG, including 
\textcolor{black}{$\lvert \mathcal{E}\rvert\cdot N$ decision variables, $3\lvert F\rvert+(2\lvert F\rvert\cdot\lvert\mathcal{V}\rvert+\lvert F\rvert\cdot\lvert \mathcal{E}\rvert)\cdot N + \lvert \mathcal{E}\rvert\cdot N$ constraints in our proposed model, where $\lvert \mathcal{V}\rvert$ is the number of nodes, $\lvert \mathcal{E}\rvert$ is the number of edges, and $\lvert F\rvert$ is the number of flows. However, the model presented in \cite{ref10} has $\lvert \mathcal{E}\rvert\cdot \lvert F\rvert^{2}$ variables and $2\lvert F\rvert+3\lvert F\rvert\cdot \lvert \mathcal{V}\rvert+\lvert F\rvert\cdot\lvert \mathcal{E}\rvert+\lvert F\rvert^{2}\cdot\lvert\mathcal{E}\rvert$constraints.
Clearly, TSEG allows us to model the problem TSEG-MAX, reducing $\lvert \mathcal{E}\rvert\cdot(\lvert F\rvert^{2}-N)$ variables and around 
$(|F|+1-N)|F||\mathcal{E}|$
constraints with $|F|>N$ as compared to the standard ILP model in \cite{ref10}.}
However, TSEG-MAX is clearly non-convex due to its integer variables, using brute force method\cite{wangpengglobecom} requires a complexity of {$O(\lvert RK2^{R}\rvert)$, which is intractable, while $R$ and $K$ are the number of variables and constraints, respectively}. To solve TSEG-MAX efficiently, we design a TSEG-based graph algorithm in the following Section, which is proved to have good schedule quality and be fast in the Simulation Section. 

\section{Online Joint Routing and Scheduling Scheme}\label{Time-Varying Graph-Based Algorithm}
In this section, we first propose a dynamic weighting method to reveal the possibility of accommodating different flows for each edge in TSEG. Next, based on the weighting method and time-slot expanded graph, we design a joint routing and scheduling algorithm for each arrival flow, such that the effect on future resource assignment brought by allocating resources for the current flow is minimized.  

\subsection{Dynamic Weighting Method}
The key motivation for us to design a dynamic weighting method is the potential conflict relationship among flows with different periods.  
Let us recall the conflict relationship as shown in Fig. \ref{insight}(a). Suppose we already assign resources for $f_1$ as shown in Fig. \ref{insight}(b) or (c). Before $f_2$ joins the network, the considered TTEthernet can potentially deliver two types of flow, i.e., a flow with a period of 2 time slots along path $(s,b,d)$ and a flow with a period of 4 time slots along path $(s,b,d)$ or path $(s,a,d)$. Obviously, the flow with a 4-time-slots period and the flow with a 2-time-slots period are in conflict on path $(s,b,d)$, more specifically, if we arrange a flow with a period of 4 time-slots over path $(s,b,d)$, TTEthernet no longer supports the transmission of the flow with a 2-time-slots period. However, if we transmit the flow with a 4-time-slots period over path $(s,a,d)$, the 2-time-slots period flow can still be supported in TTEthernet, i.e., over path $(s,b,d)$. 

As a result, when deciding which edge (representing the available time slot of a communication link) in TSEG to transmit a frame,  with different edges supporting to transmit different periods of flows, choosing the edge supporting a larger period is always a better choice. The heuristic is that the flow with a larger period requires fewer resources as compared to the flow with a smaller period, thus having more chances to be successfully scheduled than a smaller period flow. However, if we use the edge which can transmit a smaller period flow to transmit a larger period flow, the whole network could lose a chance to transmit smaller period flows, making it more difficult to accommodate smaller period flows in TTEthernet.     

Based on this observation, we design a dynamic weighting method to reveal which kind of flows (i.e., flow with what period) one edge in TSEG can potentially transmit. Actually, we can decide this via the following constraint (\ref{eq:7}). Specifically, an edge $(u_i,v_i)$ of TSEG can transmit a flow $f$ with a period of $p_f$ time slots, iff link $(u,v)$ is available in $\tau_i$ and becomes available every $p_f$ time slots (e.g., edge $(u_{i+p_f},v_{i+p_f})$ exits) within $T$. Using such a method, we can identify the periods of potentially supporting flows for each edge in TSEG. We denote edge $(i_q,j_q)$'s potentially supported flows' period set as  $\mathcal{P}_{i,j,q}$. Based on the set, we formally define a weight $w_{i,j,q}$ for each transmission edge  $(i_q,j_q)$ in TSEG to reflect that edge's transmission capability, 
\begin{equation}\label{equ:weight}
w_{i,j,q}=\sum_{p\in\mathcal{P}_{i,j,q}}{\alpha^{\frac{N}{p}}}, \forall (i_q,j_q)\in \mathcal{E}'.
\end{equation}
here, $\alpha$ is any positive integer no less than 2. We use the number of times $\frac{N}{p}$ a flow with period $p$ needs to be transmitted within $T$ to calculate the weight. Obviously, the smaller the period of the supported flow is, the greater value it contributes to the total weight. Furthermore, one edge with a greater weight will strictly support a smaller period flow than an edge with a smaller weight.  In particular, since the capacity of caching edge is infinite, we set the weight of each caching edge as $0$.
In the online TTEthernet environment, new flow request keeps arriving. Every time we successfully schedule a flow in TTEthernet, some edges' weights and supported period set in TSEG will change. Specifically, suppose we already calculate each transmission edge's supported period set and weight in TSEG. Now let us illustrate how to update after we successfully arrange a new flow $f$ with period $p_f$ to be sent over transmission edges set $\mathcal{E}'_f$ with $\mathcal{E}'_f \subset\mathcal{E}$.  
For each edge $(i_q,j_q)$ in $\mathcal{E}'_f$, regarding any supported period $p$ in $\mathcal{P}_{i,j,q}$ except $p_f$, edge $(i_q,j_q)$ supports to transmit the flow with period $p$ together with other edges $(i_{q+kp},j_{q+kp})$ with $k\in \{k\in\mathbb{Z}:q+kp\in [1,N],k\neq 0\}$. Since we need to remove edge $(i_q,j_q)$ in TSEG, these edges can no longer transmit the flow with period $p$. Thus, edge $(i_{q+kp},j_{q+kp})$'s supported period set no longer includes $p$, with  $\{q+kp\in [1,N]|k\in Z\}$. After updating each affected edge's supported period set, we can update the weights according to equation (\ref{equ:weight}). 
For completeness, we formally give the dynamic weighting algorithm as shown in Algorithm \ref{alg:weighting}.

\begin{algorithm}[th]\caption{Dynamic weighting algorithm for TSEG.}\label{alg:weighting}
\begin{algorithmic}[1]
\STATE{\textbf{Input:} TSEG $G=\{\mathcal{V}',\mathcal{E}'\}$, flow $f$'s traversed transmission edges set $\mathcal{E}'_f$, $\mathcal{P}_{i,j,q}$ and $w_{i,j,q}$ for $\forall (i_q,j_q)\in \mathcal{E}'$}.
\STATE{\textbf{Output:} updated $\mathcal{P}_{i,j,q}$ and $w_{i,j,q}$ for $\forall (i_q,j_q)\in \mathcal{E}'$.}
\STATE{\textbf{Initialize} an edge set $\mathcal{E}''=null$ to collect effected edges.}
\FOR{$\forall (i_q,j_q) \in \mathcal{E}'_f$}
\FOR{$\forall p \in \mathcal{P}_{i,j,q}$ except $p_f$}
\FOR{$\forall k \in \{k\in \mathbb{Z}:q+kp\in[1,N],k\neq 0\}$}
\STATE{\textbf{Remove} $p$ from $\mathcal{P}_{i,j,q+kp}$}
\STATE{\textbf{Add} edge $(i_{q+kp},j_{q+kp})$ into $\mathcal{E}''$.}
\ENDFOR
\ENDFOR
\ENDFOR
\FOR{$\forall (i_q,j_q)\in \mathcal{E}''$}
\STATE{\textbf{Update} $w_{i,j,q}$ according to (\ref{equ:weight}).}
\ENDFOR
\end{algorithmic}
\end{algorithm}

\subsection{Interesting Observations for Weights}
\underline{\textit{Observation 1:}} In general, the total weights of all edges in TSEG indicate how many flows TTEthernet can potentially accommodate. 

\underline{Proof:}  Let us consider the case when no flows join TTEthernet,  TTEthernet has the most available resources. At this time, according to the formula (\ref{equ:weight}), the corresponding TSEG has the largest total weight of all edges. As TTEthernet starts to accommodate new flows, according to Algorithm \ref{alg:weighting}, the weights of some edges strictly decrease while the weights of all the other edges keep unchanged, hence the total weights of edges strictly decrease with more flows joining the network. As a result, accommodating more flows requires a larger total weight of all edges in TSEG. On the other hand, the weight of edges does not indicate how many frames can be transmitted over the edges, actually, one edge can at most transmit one frame. The weight indicates which kind of flow can be supported. Specifically, the larger the weight of an edge is, the smaller the period of flow it can support. Thus, the large total weight of TSEG edges can represent the ability of TTEthernet to accommodate small-period potential flow. Since we cannot know which kind of flow will arrive, a large total weight will indicate a greater chance to accommodate a potential new flow.      

\underline{\textit{Observation 2:}} The total weights of edges in TSEG strictly decrease every time a new flow is scheduled. 

This property is obvious according to Algorithm \ref{alg:weighting}. 

\underline{\textit{Observation 3:}} Reduction of the sum of all edges' weights, incurred by assigning resources to a new flow, represents the effect of current resource assignment on future potential flows.

\underline{Proof:} 
As declared by \textit{Observation 2}, successfully scheduling a flow in TTEthernet will necessarily reduce the sum of all edges' weights. On this premise, we further identify when we try to find paths in TSEG to accommodate a newly arrived flow, choosing different paths could result in different weight reduction, making different effects on the resource allocation for future flow. For instance, suppose a flow with a period of $p$ time slots arrives, there exist three paths in TSEG named $path_1$, $path_2$, and $path_3$. We assume $path_1$ can support to transmit the flow with period of $p$ time slots and the flow with $p_1$ time slots period, with $p_1<p$ (i.e., each edge along $path_1$ supports to transmit flows with period of $p$ and $p_1$ time slots), while the smallest period of flow $path_2$ and $path_3$ support to transmit is $p$, besides, we let $path_2$ contain some edges which support to transmit flows with a $p_1$-time-slots period. 
If we choose $path_1$ to transmit a frame of the flow, TTEthernet will permanently lose a chance to transmit a flow with a $p_1$-time-slots period. If we choose $path_2$, since some edges of $path_2$ support transmitting flows with a $p_1$-time-slots period, the occupation of entire $path_2$ will necessarily disable some edges in TSEG to transmit $p_1$-time-slots period flow, the affected number of edges will be smaller than choosing $path_1$. In comparison, if we choose $path_3$, the effect will be the minimum. Correspondingly, according to Algorithm \ref{alg:weighting}, the weights reduction will be the highest by choosing $path_1$, while the smallest reduction can be achieved via choosing $path_3$. 

\underline{\textit{Observation 4:}} Occupying an edge with a larger weight will incur more weight reduction as compared to occupying an edge with a smaller weight. 

\underline{Proof:} According to the formula (\ref{equ:weight}), the larger weight of an edge indicates the ability to transmit smaller period flows.  Recall that when an edge $(u,v)$ supports transmitting a $p$-time-slots period flow in TSEG,  other $\frac{N}{p}-1$ edges support transmitting such a flow. When $(u,v)$ is occupied to transmit other flow, those  $\frac{N}{p}-1$ edges no longer support transmitting the $p$-time-slots period flow, the corresponding weights should be reduced from these  $\frac{N}{p}-1$ edges. As a result, when a larger weight edge is occupied, the weight of more edges will be affected and the weight reduction for each affected edge is larger as per formula (\ref{equ:weight}). As a result, the observation is proved.     

With the above observations, we remark that to minimize the effect of current flow on future flow, choosing a path with a minimum total weight among all the paths is necessary. These observations justify why we design a minimum weight path algorithm in the next subsection.

\subsection{Flows-friendly Joint Routing and Scheduling Algorithm}

Recall that our considered TSEG-MAX aims at maximizing the successfully scheduled number of flows with the arrival of flows unaware. To schedule as many arrived flows as possible, one promising way is to reduce the effects of arranging the current flow on the resource assignment for upcoming flows. According to the \textit{Observations} in the previous subsection, we design a flows-friendly joint routing and scheduling algorithm which finds the  \textit{path with minimum weights} (here the path weights refer to the total weights of path's all edges) in TSEG as shown in Algorithm \ref{alg:alg1}, reducing the effects on future flows while satisfying the requirements of current flows. In particular, finding the path with minimum weights is essentially a shortest path problem, existing shortest path algorithms (e.g., Dijkstra algorithm\cite{dijkstra}, Bellman-Ford algorithm\cite{bellman}, A$^*$ search algorithm \cite{A*}), however, cannot apply, due to the unique flow conservation constraints (\ref{eq:input}) and (\ref{eq:output}), and unique QoS constraints (\ref{eq:7}) and (\ref{eq:e2e}).



To find the minimum-weight path while satisfying those special constraints, we modify the Dijkstra algorithm to propose Algorithm \ref{alg:alg1}. The subroutine is similar to that of the Dijkstra algorithm. That is, we maintain a queue $Q$ collecting all the vertices in TSEG initially. Each vertex in $Q$ is associated with 2 attributes $\mathcal{H}(\cdot)$ and $\mathcal{R}(\cdot)$, with $\mathcal{H}(\cdot)$ representing the smallest weight of the path to source node $s$ in current iteration, while  $\mathcal{R}(\cdot)$ representing \textcolor{black}{second} hop node of the shortest path to node $s$. Every time we take out the vertex $u_q$ with smallest $\mathcal{H}(\cdot)$ in $Q$ and update $\mathcal{H}(\cdot)$ and $\mathcal{R}(\cdot)$ of $u_q$'s certain neighbor vertices if their weights satisfying the relaxation condition as in \cite{dijkstra}. We repeat the above operations until the representation vertices of destination node $d$ are taken out in $Q$. Despite the similarity, we make the following major modifications in Algorithm \ref{alg:alg1} to meet the special constraints for TTEthernet flows:

a): To ensure periodical transmission as indicated by (\ref{eq:7}), after we take out $u_q$ with smallest $\mathcal{H}(\cdot)$ from $Q$ in each iteration, we only consider the outgoing edges $(u_q,v_q)$ of $u_q$ which supported to transmit the flow with period $p$, which is different from the Dijkstra algorithm considering each neighbor vertices of $u_q$. Specifically, we require the supported periods set $\mathcal{P}_{u,v,q}$ of the outgoing edges to include period $p$ as indicated by steps 22-27. Thus, constraints (\ref{eq:7}) are satisfied.

b): To ensure the delay of flow falls within the maximum tolerable delay as indicated by (\ref{eq:e2e}), we let the algorithm find paths within the delay range. Specifically, given a hyper-period $T=[\tau_1,\tau_N]$, to successfully schedule a flow with period $p$, the flow must start to transmit within $\tau_i$ with $i\in[1,p]$, otherwise the flow should be rejected. Hence, we consider starting to transmit flow in $\tau_i$ as shown in step 3, each edge should be within $[\tau_i, \tau_{i+m-1}]$, such that the delay of the end-to-end path will not exceed $m$ time slots.
Hence, we require the neighboring vertices of $u_q$ to lie in a time slot $\tau_l$ with $l\leq i+m-1$ as shown in steps 22-27.

c): To satisfy the unique flow conservation (\ref{eq:input}) and (\ref{eq:output}), let us recall these two constraints first. For any node $u$, if the flow enters it through a transmission edge, the flow must leave it through a caching edge as indicated by (\ref{eq:input}). On the other hand, there is no restriction on the choice of outgoing edges if the flow enters $u$ through a caching edge as indicated by (\ref{eq:output}). These two constraints imply that the vertex sending data to $u$ and the vertex receiving data from $u$ must represent nodes of different time slots. Hence, for source vertex $s_i$, we set its parent vertex $s_{i-1}$ to indicate no constraints on the choice of the outgoing edges for the source vertex. Next, for each vertex $u_q$, we constrain its parent vertex $i_k$ and next hop vertex $v_l$ to represent nodes of different time slots via constraining $l\neq k$ as indicated by steps 20-27. 

Note that, we cannot know the minimum-weight path should start from which time slot $\tau_i$ with $i\in[1,p]$ in advance, we record the minimum weight path starting from each $i\in[1,p]$ in the path set $\mathcal{O}$, and finally output the path in $\mathcal{O}$ with a minimum weight. Every time we schedule a flow, we should invoke algorithm \ref{alg:weighting} to update the supported period set and weights of each edge in TSEG.

\begin{algorithm}[th]\caption{Flows-friendly joint routing and scheduling algorithm (JRAS-TSEG).}\label{alg:alg1}
\begin{algorithmic}[1]
\STATE{\textbf{Input:} TSEG $=\{\mathcal{V}',\mathcal{E}'\}$,  flow  $f=\{t,s,d,p,m\}$, $\mathcal{P}_{i,j,q}$ with $ \forall (i_q,j_q) \in \mathcal{E}'$}.
\STATE{\textbf{Output:} flow $f$'s path if successfully scheduled. }
\FOR{$\forall i\in [1,p]$}
\STATE{\textbf{Initialize} a path set $\mathcal{O}=\{\}$}
\STATE{\textbf{Initialize} a queue $Q=\{\}$.}
\FOR{$\forall u_q\in \mathcal{V}'$}
\STATE{ $\mathcal{R}(u_q)=null$.}
\STATE{ $\mathcal{H}(u_q)=+\infty$.}
\STATE{$Q\leftarrow u_q$.}
\ENDFOR
\STATE{ $\mathcal{H}(s_i)=0$.}
\STATE{ $\mathcal{R}(s_i)=s_{i-1}$.}
\WHILE{$Q\neq \emptyset$}
\STATE{\textbf{Find} $u_q= \arg\min\{\mathcal{H}(i_q)|i_q\in {Q}\}$.}
\IF{$u==d$}
\STATE{\textbf{Add} identified path to $\mathcal{O}$.}
\STATE{\textbf{Break}.}
\ENDIF
\STATE{$Q\rightarrow{u_q}$.}
\STATE{ $i_k=\mathcal{R}(u_q)$.}
\FOR{each node $v_l\in {Q}$ adjacent to node $u_q$}
\IF{$k$==$l$ \OR $l>i+m-1$ \OR $p\notin \mathcal{P}_{u,v,q}$}
\STATE{\textbf{Continue}}
\ELSIF{$\mathcal{H}(v_l) \geq \mathcal{H}(u_q)+w_{u,v,q}$}
\STATE{ $\mathcal{H}(v_l)=\mathcal{H}(u_q)+w_{u,v,q}$.}
\STATE{ $\mathcal{R}(v_l)=u_q$.}
\ENDIF
\ENDFOR
\ENDWHILE
\ENDFOR
\IF{$\mathcal{O}==\emptyset$}
\STATE{\textbf{Output:} No feasible path!}
\ELSE
\STATE{\textbf{Output:} The path in $\mathcal{O}$ with minimum weight.}
\ENDIF
\end{algorithmic}
\end{algorithm}

\subsection{Running example}
\begin{figure}[!htbp]
	\centering
	\includegraphics[width=0.5\textwidth]{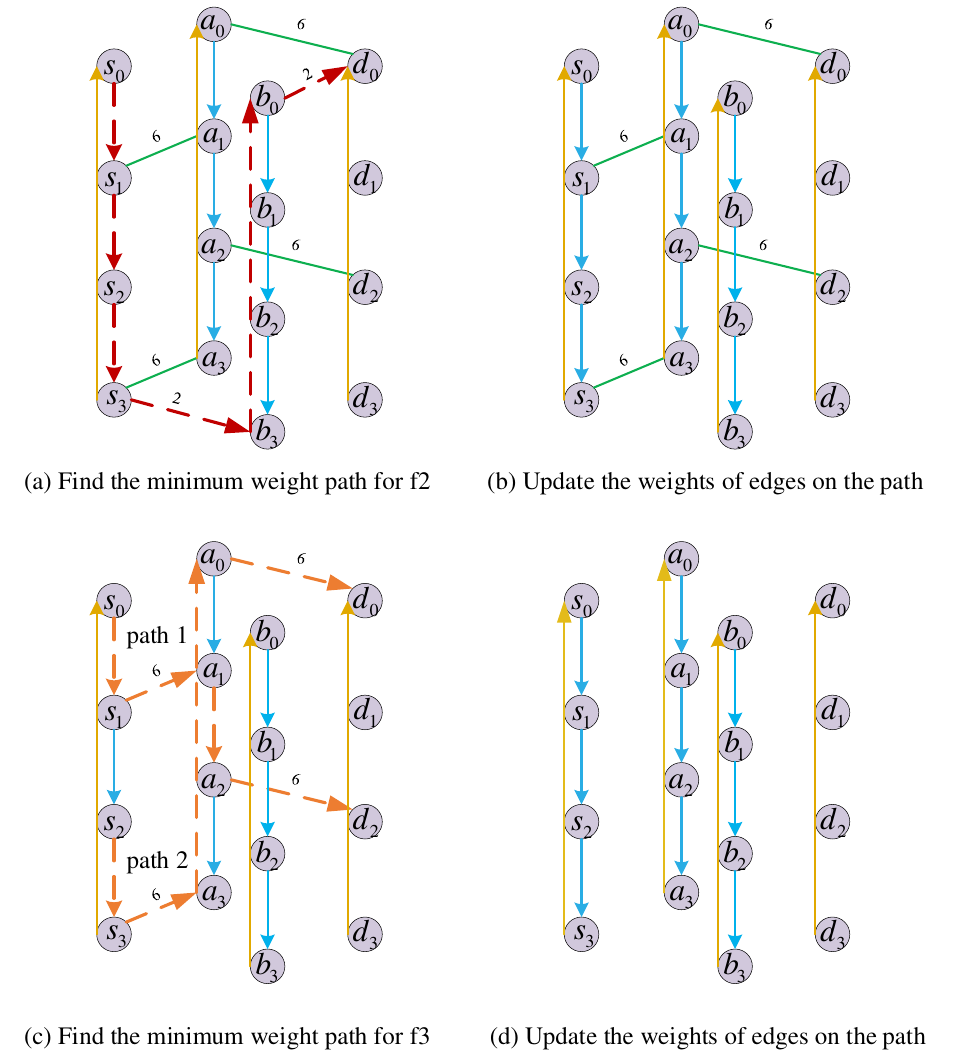} 
	\caption{Example of the JRS-TSEG algorithm, where the input is two flows with periods of 4 time slots and 2 time slots, and the output is a path represented by dashed lines.} 
	\label{algorithm example}
\end{figure}

To better illustrate the advantages of our proposed algorithm, an example describing the algorithm's solving process will be given, as shown in Fig. \ref{algorithm example}. It uses the network topology in Fig. \ref{insight}(b), where the new flow $f_1$ is already arranged on the path $(s,b,d)$.
Recall that the next two flows arriving are $f_2$ and $f_3$, with periods of 4 and 2 time slots respectively. The specific description of flow $f_2$ and $f_3$ is as follows: $f_1=\left(t_1,s,d,4,8\right)$, $f_3=\left(t_1,s,d,2,4\right)$. Let's assume that the value of $\alpha$ is 2. Fig. \ref{algorithm example}(a) shows the TSEG corresponding to the network topology, where the value on the link represents transmission capability after the network resource is updated.
The shortest path $\left(s_0,s_1,s_2,s_3,b_3,b_0,d_0\right)$ for flow $f_2$ could be found in TESG via algorithm \ref{alg:alg1}, where all edges are identified in red.
The value of the transmission edge on this path is then updated according to algorithm \ref{alg:weighting}, as shown in Fig. \ref{algorithm example}(b). Fig. \ref{algorithm example}(c) shows the path construction result of flow $f_3$. Since its period is 2 time slots, it is necessary to find the two shortest paths that meet the demand of flow transmission in the graph, and their time slot interval is the period of flow. The result of the network resource update is shown in Fig. \ref{algorithm example}(d). Obviously, flow $f_2$ with a 4-time-slots period and flow $f_3$ with a 2-time-slots period are in conflict on path $(s,a,d)$.
If the path $\left(s_0,s_1,a_1,a_2,d_2\right)$, which is the earliest path to destination node $d$, is found for flow $f_2$ and is also a path with low congestion, the remaining resources cannot meet the transmission requirements of flow $f_3$. This shows that the co-optimization of routing and scheduling based on TESG will enhance the probability of eliminating potential conflicts among flow requests.

\subsection{Complexity Analysis}
The complexity of the algorithm mainly depends on the $p$-times path construction process, where $p$ is the period of new arriving flow $f$. For each path construction process, $\lvert\mathcal{V}'\rvert$ iterations are required in the worst case. 
In each iteration, it is necessary to determine whether the transmission edge meets periodic flow transmission in the hyper-period. The time complexity of path construction process is $O(\lvert\mathcal{V}'\rvert\cdot[log_{}{\lvert\mathcal{V}'\rvert}+B\cdot(log_{}{\lvert\mathcal{V}'\rvert}+N)])$,
where $\lvert\mathcal{V}'\rvert$ is the number of nodes of TSEG, $B$ is the maximum number of neighbors of a node $u$.
$N$ is the number of time slots of hyper-period.
Since $\lvert\mathcal{V}'\rvert\cdot B=\lvert\mathcal{E}'\rvert$, the complexity of this part can be expressed as $O((\lvert\mathcal{V}'\rvert+\lvert\mathcal{E}'\rvert)\cdot log_{}{\lvert\mathcal{V}'\rvert}+\lvert\mathcal{E}'\rvert\cdot N)$, where $\lvert\mathcal{E}'\rvert$ is the number of edges of TSEG.
Thus, the worst-case time complexity of our algorithm is $O(p \cdot((\lvert\mathcal{V}'\rvert+\lvert\mathcal{E}'\rvert)\cdot log_{}{\lvert\mathcal{V}'\rvert}+\lvert\mathcal{E}'\rvert\cdot N))$, which can be reduced to $O(p \cdot \lvert\mathcal{E}'\rvert \cdot(N+log_{}{\lvert\mathcal{V}'\rvert}))$.

\section{SIMULATION RESULTS}\label{SIMULATION RESULTS}

In this section, we first introduce the setup of our 
numerical experiment. Next, we evaluate the the number of successfully scheduled flows and the running time of our proposed scheme and existing methods. The result shows that our scheme obviously outperforms the existing schemes. 

\subsection{Experiment Setup}
\subsubsection{Topology of TTEthernet}  
\begin{figure}[!htbp]
	\centering
	\includegraphics[width=0.4\textwidth]{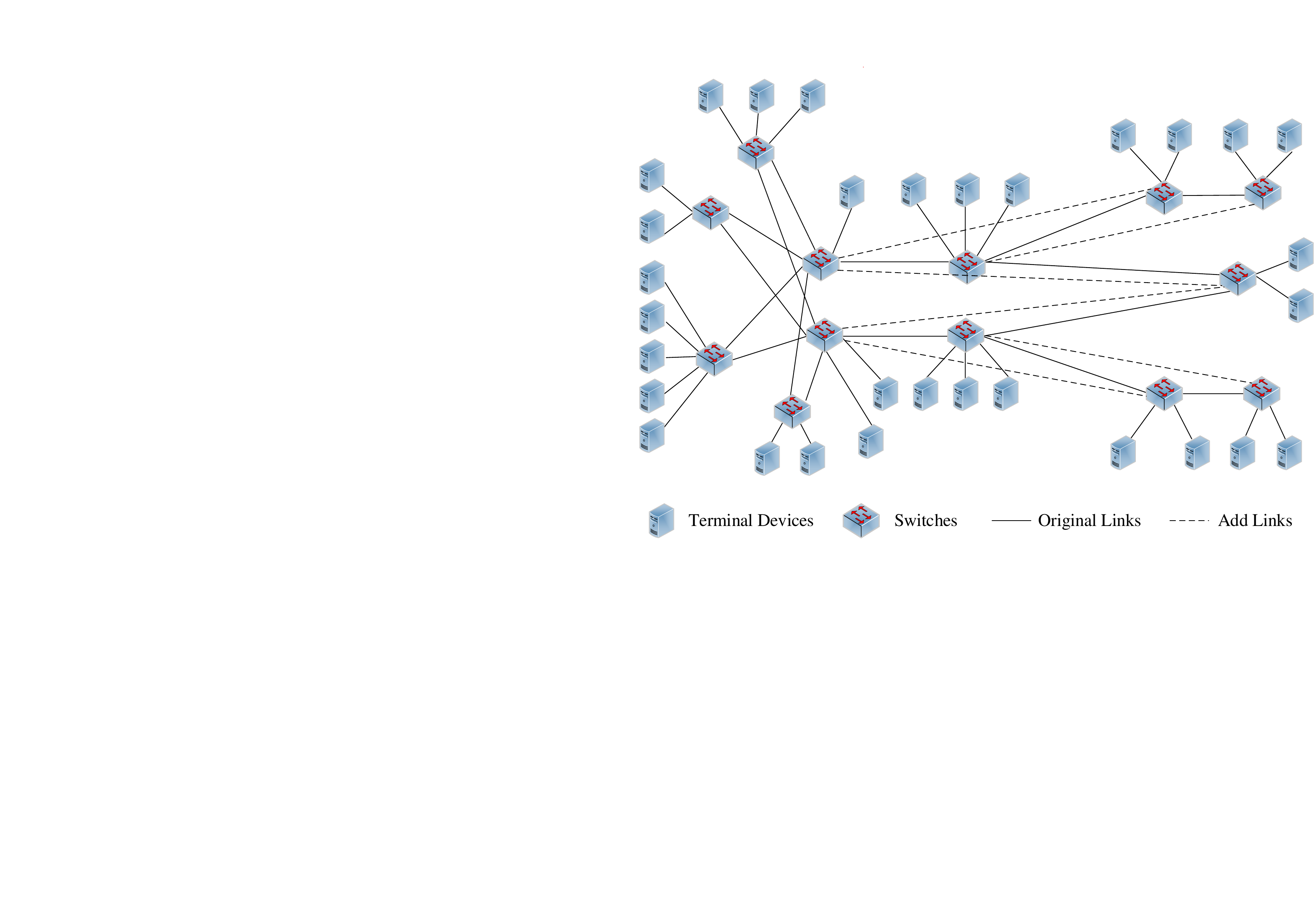} 
	\caption{CEV network topology connecting 31 terminal devices and 13 switches.} 
	\label{cevtopo}
\end{figure}

The existing industrial control networks can be categorized into star, tree, line, snowflake, and ring networks according to their structures. As a result, we use the following two typical structures as the network topology in the experiments:
\begin{itemize}
\item Ring-structure topology as in \cite{ref16}. The ring network always has two paths for any given source and destination nodes, known for their robustness against the failures of switches or communication links. 
\item Orion Crew Exploration Vehicle (CEV) network with added links as in \cite{ref39}. The CEV network can be viewed as a combination of 5 typical structures as shown in Fig. \ref{cevtopo}. It consists of 13 switches and 31 terminal systems.
\end{itemize}

We set the data transmission rate for the full-duplex link in TTEthernet as 1 Gbit/sec as in \cite{ref10}. Hence, the length of a time slot is set to $12\mu s$, and the propagation delay is neglected. 


%
%

\subsubsection{Setup for Time-triggered Flows} We first introduce how to generate each flow. In the experiments, all flows are generated randomly according to the guidance of the traffic characteristics defined in the IEC/IEEE 60802 standard\cite{ref35}. \textcolor{black}{Regarding the ring network, we randomly select nodes from the whole network as source and destination nodes for each flow. While in the CEV network, we randomly choose nodes only from the end systems to be the flow's source and destination.} 
In addition, we randomly select a value from the set $\{60, 120, 240, 480\}$ (unit: $\mu s$)  or the set $\{60, 480\}$ (unit: $\mu s$) as the length of flow period.  
Each flow's maximum allowed delay is set to four times its period. 

Now let us define the pattern of all the considered flows according to the period of flows. We use the period of flow to define the pattern due to the observation that flows with larger transmission periods consume fewer resources than the larger-period flow and are thus easier to be successfully scheduled.  As such,
we define the ratio of the flows with a 60 $\mu s$-period to total flows as $\alpha$. Similarly, we define the ratio for $120 \mu s$-period flows, $240 \mu s$-period flows and $480 \mu s$-period flows as $\beta$, $\lambda$ and $\delta$, respectively, as shown in Table \ref{Traffic Classes}. Naturally, if the  periods of flows are chosen from $\{60, 120, 240, 480\}$, there is $\alpha+\beta+\lambda+\delta=1$, otherwise, there is $\alpha+\lambda=1$ for period set $\{60, 480\}$ (unit: $\mu s$). In our considered online schedule environment, we allow at most one flow can arrive at the network at the same time. 

\begin{table}[htbp]
	\centering
	\caption{Traffic Classes.}
	\label{Traffic Classes}
	\resizebox{0.4\textwidth}{!}{
	\begin{tabular}{| c | c | c | c |}
		\hline
		Class & Period ($\mu s$) & Max delay ($\mu s$) & Flowcount\\
		\hline
		1& 60 & 240 & $\alpha\ast \lvert F\rvert$ \\
		\hline
		2& 120 & 480 & $\beta\ast \lvert F\rvert$ \\ 
		\hline
		3& 240 & 960 & $\lambda\ast \lvert F\rvert$ \\
		\hline
		4& 480 & 1520 & $\delta\ast \lvert F\rvert$ \\
		\hline 
	\end{tabular}
}
\end{table}


\subsubsection{Baselines} To demonstrate the performance gain of our approach in terms of runtime and scheduling quality, we consider the following 4 different schemes as baselines, including one offline scheme and three online schemes:
\begin{itemize}
\item \textit{Optimal Solution}, which is implemented via using the solver Groubi\cite{ref36} to solve the ILP problem formulated in section \ref{System Model}. \textit{Optimal solution} is an offline method, able to calculate the maximum number of successfully scheduled flows. It can give an achievable upper bound to evaluate the optimality gap of our scheme. 
\item  \textit{IRAS} as in \cite{ref6}, which first finds all the paths between source and destination nodes, and weighs each path according to the number of hops and the occupied number of time slots of each link along the path. Next, IRAS selects the smallest-weight path to check whether the flow can be transmitted along the chosen path by solving an ILP problem. If not successfully scheduled, IRAS chooses the smallest-weight path of the remaining paths to check whether the flow can be accommodated. The above operation is repeatedly performed until IRAS finds a path that can accommodate the flow. To implement this baseline, we calculate paths through Networkx library\cite{ref37} and solve the ILP problem using the Gurobi solver\cite{ref36}.
 \item \textit{SRS-TSEG}.  We propose \textit{SRS-TSEG} using the idea of weighted time slots when allocating paths for flows as in \cite{ref7}. The main
 difference between \textit{SRS-TSEG} and \textit{IRAS} is \textit{SRS-TSEG} weights the time slots over each link aiming at accommodating more flows in TTEthernet. To implement \textit{SRS-TSEG}, we have modified our proposed \textit{JRS-TSEG} to independently find a path for a flow and assign weighted time slots to a flow.   
 \item \textit{JRS-TESG (wop)}. We further propose a variation of our \textit{JRS-TSEG} as \textit{JRS-TESG (wop)}. Recall that \textit{JRS-TSEG} schedules the flows by finding the smallest weight path. Here,  \textit{JRS-TESG (wop)} simply finds a path with the smallest hops in TSEG (i.e., smallest delay), without considering the weights of links. We propose \textit{JRS-TESG (wop)} to verify the effectiveness of our weighting method. 
\end{itemize}

All experiments are run on an Intel(R) Core(TM) i7-10700 64-bit CPU @ 2.90GHz with 16 GB of RAM.

\subsection{Performance Evaluation under Different Topologies}
In this experiment, we evaluate the performance of the proposed scheme and the other four baselines under different sizes of network topology. Note that, we can only evaluate the performance of \textit{Optimal solution} under small-size networks due to the high complexity of solving an ILP. 
We randomly choose a number from set $\{60, 120, 240, 480\}$ (unit: $\mu s$) as a flow's period. We set the flow pattern as $\alpha=0.2$, $\beta=0.2$, $\lambda=0.3$ and $\delta=0.3$. Under each topology, we run the experiment ten times to obtain an average performance.

\subsubsection{Evaluation of successfully scheduled number of flows} 

\begin{figure*}
	\begin{minipage}{0.33\textwidth}
		\centering
		\includegraphics[width=0.9\linewidth]{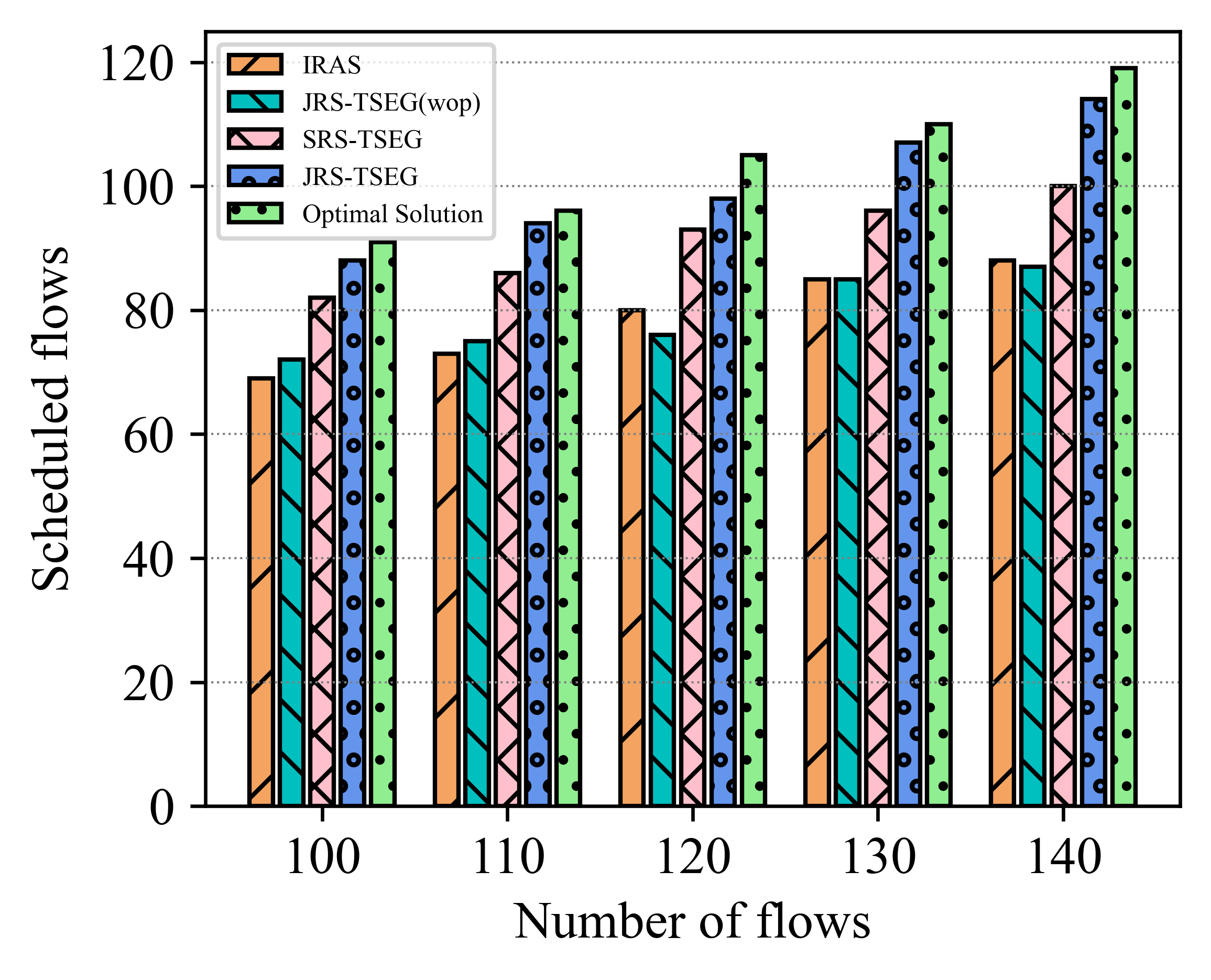}
		\caption{The comparison of scheduled flows among different algorithms in the ring topology with period set $\{60, 120, 240, 480\}$ and $\alpha=0.2$, $\beta=0.2$, $\lambda=0.3$, $\delta=0.3$.}
		\label{fz1}
	\end{minipage}
	\begin{minipage}{0.33\textwidth}
		\centering
		\includegraphics[width=0.9\linewidth]{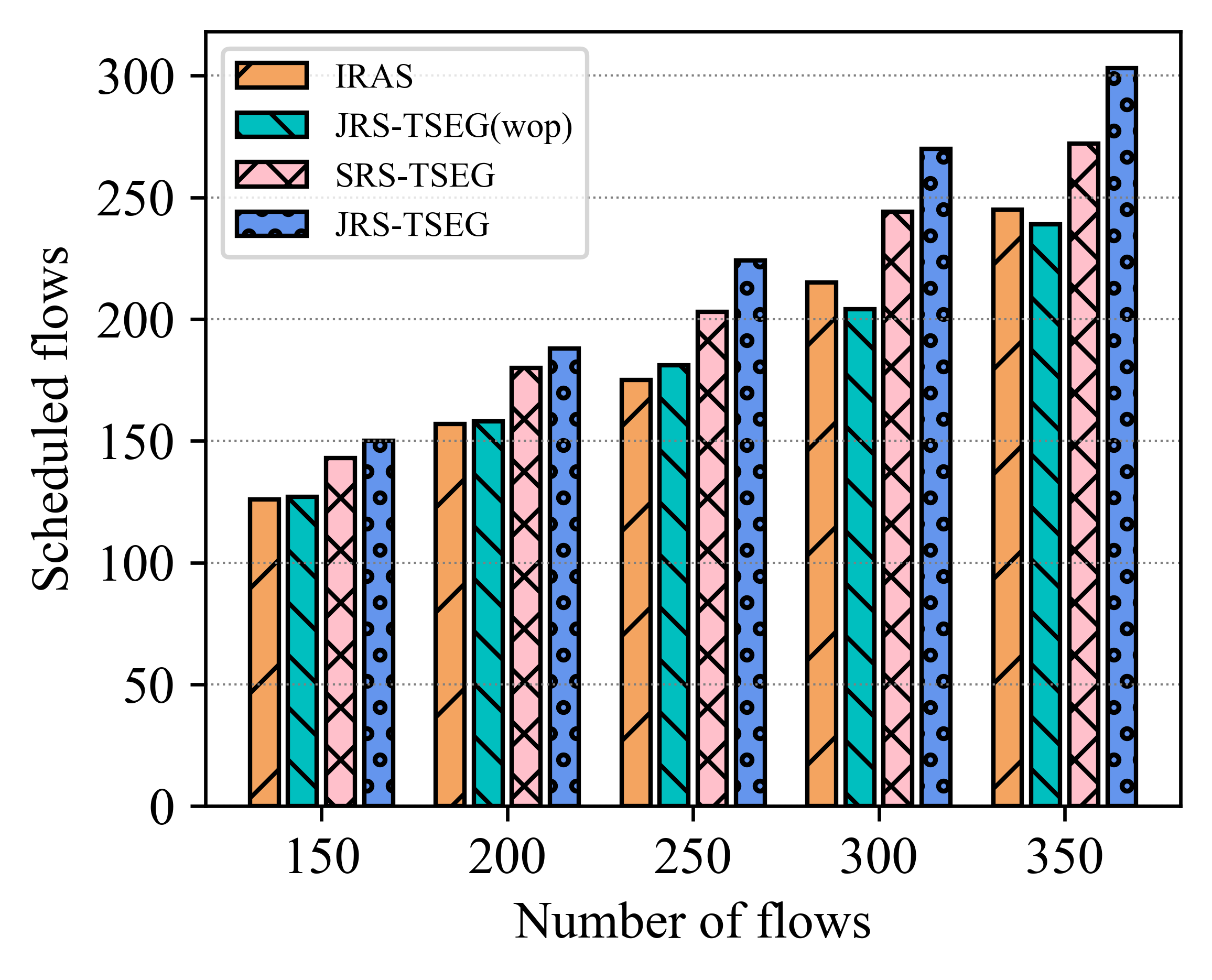}
		\caption{The comparison of scheduled flows among different algorithms in the Orion CEV network with period set $\{60, 120, 240, 480\}$ and $\alpha=0.2$, $\beta=0.2$, $\lambda=0.3$, $\delta=0.3$.}
		\label{fz2}
	\end{minipage}
	\begin{minipage}{0.33\textwidth}
		\centering
		\includegraphics[width=0.9\linewidth]{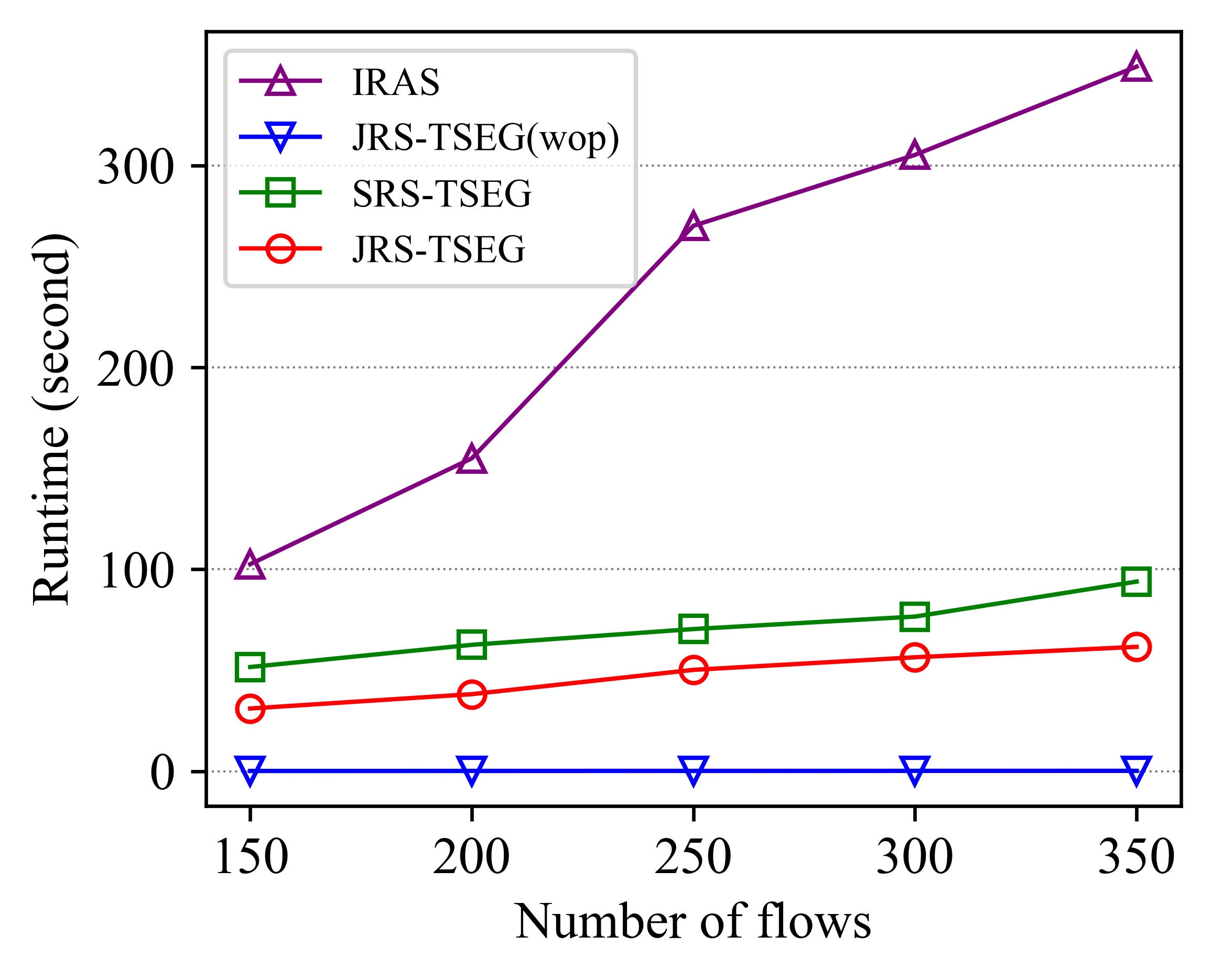}
		\caption{The comparison of runtime among different algorithms in the Orion CEV network with period set $\{60, 120, 240, 480\}$ and $\alpha=0.2$, $\beta=0.2$, $\lambda=0.3$, $\delta=0.3$.}
		\label{fz4}
	\end{minipage}
\end{figure*}

Under a ring topology with $12$ nodes, we evaluate the successfully scheduled number of flows of the 5 schemes with the number of given flows increasing from 100 to 140 with a step of 10 as shown in  Fig. \ref{fz1}. The number of successfully completed flows computed by our algorithm reaches an average of 98\% of that of the optimal solution. 
Meanwhile, the numbers of successfully scheduled flows by IRAS, JRS-TSEG(wop), and SRS-TSEG reach 75\%, 75\%, and 86\% of the number of optimally scheduled flows, respectively. \textit{SRS-TSEG} outperforms \textit{IRAS} and \textit{JRS-TSEG(wop)}  
due to the employed dynamic weighting method. Specifically, the weighting method allows to assign the time slots supporting large-period flows to the current flow instead of choosing the time slots supporting both large-period and small-period flows, protecting the ability of the network to transmit small-period flows. 
In contrast, IRAS and JRS-TSEG(wop) focus on selecting a proper path and then deciding when to transmit a packet on each link along the chosen path to deliver the packet to the destination as early as possible. As a result, the time slots supporting small-period flows can be allocated to large-period flows. Furthermore, our scheme outperforms \textit{SRS-TSEG} due to the fact \textit{SRS-TSEG} finds a path first and then decides when to transmit data over each link while our scheme co-optimizes the routing and scheduling. Specifically, rather than fixing a path first and then occupying resources of the path, our scheme integrates path-finding and scheduling into a single process, allowing us to find a more proper path with more suitable time slots available. 




Fig. \ref{fz2} evaluates the schedulability of the 4 schemes under the CEV topology. Notably, we do not employ the \textit{optimal solution} here due to the large size of the CEV topology. The number of flows increases from 150 to 350 with a step of 50. 
Due to the same reason, the number of our scheme's successfully scheduled flows is 23\% larger than IRAS and JRS-TSEG (wop). Besides, the performance gain of our scheme over SRS-JSEG reaches 12\%.  This experiment proves that our scheme still performs well in large-scale environments.

We observe that the performance gain of our scheme Under the ring topology is bigger than the gain in the CEV network, while the CEV network obviously has a larger communication capacity than the ring topology since it can accommodate more flows. This is because the performance gain mainly depends on the increase in the completion number of small-period flows while large-period flows can be easily scheduled by all the schemes. For instance, when we insert more flows into networks (e.g., from 140 in the ring network to 350 in the CEV network), recalling that there is a fixed ratio among the inserted flows with different periods, the increase in the number of finished small-period flows cannot catch up with the increase in the number of finished large-period flows, resulting in a smaller performance gain. 
This comparison clearly indicates that the small-period flows can significantly affect the performance gain of our scheme, motivating us to explore the effects of different flow patterns and different period values on the performance in the following experiments.



\subsubsection{Evaluation of Runtime} 
Table \ref{ring_runtime} presents the total runtime of scheduling all the flows under a ring topology for the 5 schemes.   
Particularly, even under such a small topology, the \textit{optimal solution} still consumes hundreds even thousands of seconds to solve the problem, while the TSEG-based schemes only require around a second, indicating the strength of using a graph to explore the special structure of problem against using a general algorithm (e.g., the simplex method adopted by the Gurobi solver to solve ILP) in solving a combination optimization problem.  
In particular, our scheme only needs an average of 3 milliseconds to schedule each flow, much smaller than that of IRAS and SRS-TSEG schemes. This is as expected since IRAS and SRS-TSEG  enumerate all the paths before choosing a path while our scheme directly identifies one path. This drawback of IRAS and SRS-TSEG can be exacerbated when the network is more dense. The smallest runtime of JRS-TSEG(wop) further indicates the advantages of TSEG in the strength of finding paths using a graph against solving one ILP problem to find paths.
 

\begin{table}
	\centering
	\caption{The average running time of different algorithms}
	\label{ring_runtime}
	\begin{tabularx}{0.5\textwidth}{m{1.1cm} m{1.1cm} m{1.1cm} m{1.1cm} m{1.1cm} m{1.1cm}} 
		\toprule
		   & 100 flows & 110 flows & 120 flows & 130 flows & 140 flows \\
		\midrule
		JRS-TSEG & 0.16 s & 0.21 s & 0.31 s & 0.38 s & 0.51 s  \\
            \addlinespace 
           SRS-TSEG & 1.98 s & 2.07 s & 2.11 s & 2.19 s & 2.25 s  \\
            \addlinespace 
            JRS-TSEG(wop) & 0.009 s & 0.016 s & 0.022 s & 0.026 s & 0.028 s  \\
            \addlinespace 
		IRAS & 5.10 s & 5.74 s & 6.45 s & 6.90 s & 8.50 s  \\
            \addlinespace 
		Optimal Solution & 82.26 s & 130.75 s & 330.32 s & 845.38 s & 1761.53 s \\
		\bottomrule
	\end{tabularx}
\end{table}


Fig. \ref{fz4} shows the runtime of all the schemes except the \textit{optimal solution} under the CEV topology.
As the number of inserted flows increases, 
the runtime of all methods increases, which is obvious. In particular, 
the increase in the runtime of our scheme is very steady as compared to the sharp increase of the IRAS scheme.  Again, our scheme has good scalability. 
In addition, despite that JRS-TSEG(wop) has a smaller runtime than our scheme, the 23\% performance gain of our scheme over JRS-TSEG(wop) as shown in Fig. \ref{fz1}  makes this trade-off worthwhile.

\begin{figure}
	\begin{minipage}{0.5\textwidth}
		\centering
		\includegraphics[width=0.7\linewidth]{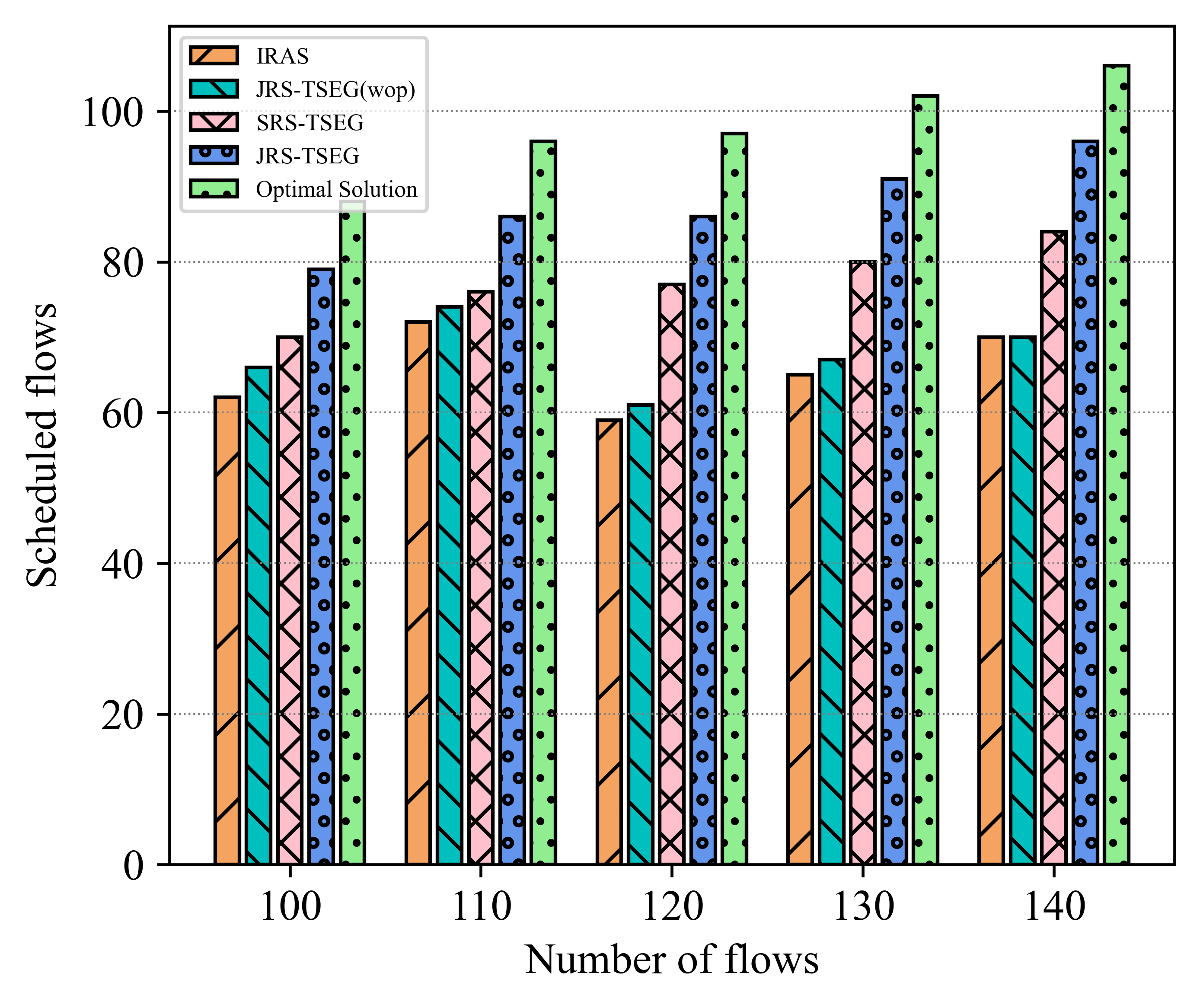}
		\caption{The comparison of scheduled flows among different algorithms in the ring topology with period set $\{60, 120, 240, 480\}$ and $\alpha=0.3$, $\beta=0.3$, $\lambda=0.2$, $\delta=0.2$.}
		\label{fz8_1}
	\end{minipage}
\end{figure}

\subsection{Effect of Different Flow Ratios On Performance}
In this section, we will explore the effect of changing flow proportions on scheduling performance. In this context, we augment the ratio of flow with periods $60 \mu s$ and $120 \mu s$, while diminishing the proportions of flow from other periods. This adjustment is motivated by the observed disparity in scheduling capabilities between our method and alternative approaches, as manifested in previous simulations.
The parameters in Fig. \ref{fz8_1} are set to ring topology, period set $\{60, 120, 240, 480\}$ (unit: $\mu s$), and the flow proportions $\alpha=0.3$, $\beta=0.3$, $\lambda=0.2$ and $\delta=0.2$ respectively.
As shown in Fig. \ref{fz8_1}, the number of scheduled flows of all methods decreased compared with the data in Fig. \ref{fz1}, because small period flows increased and needed to occupy more time slot resources.
Different from the results in Fig. \ref{fz1}, our method can schedule 33.6\% more flows than IRAS and JRS-TSEG(wop) on average and 14\% more flows than SRS-TSEG on average. This indicates that as the number of input flows increases, the number of flows affected by potential conflicts in this setting increases.

Furthermore, we give the comparison between our method and other baselines in terms of runtime under this parameter setting, as shown in Table \ref{runtime_8_2}.
Since the running time of the Optimal Solution exceeds 2 hours, the running time of the corresponding flow is null.
It shows that the runtime of our method and JRS-TSEG(wop) remain low, while the run time of IRAS and SRS-TSEG has increased significantly and maintained a continuously increasing trend.
Combining the results from the figure and table implies that our approach achieves higher resource utilization in a shorter time, that is, scheduling more flows. This is very meaningful for dynamic scenarios where flows frequently initiate requests, which not only quickly respond to the needs of flows, but also keep network resources from being idle for a long time.

\begin{table}
	\centering
	\caption{The average running time of different algorithms}
	\label{runtime_8_2}
	\begin{tabularx}{0.5\textwidth}{m{1.1cm} m{1.1cm} m{1.1cm} m{1.1cm} m{1.1cm} m{1.1cm}} 
		\toprule
		& 100 flows & 110 flows & 120 flows & 130 flows & 140 flows \\
		\midrule
		JRS-TSEG & 1.60 s & 1.31 s & 1.68 s & 1.53. s & 1.99 s  \\
             \addlinespace 
            SRS-TSEG & 7.95 s & 8.28 s & 8.87 s & 9.02 s & 9.47 s  \\
            \addlinespace 
            JRS-TSEG(wop) & 0.015 s & 0.018 s & 0.019 s & 0.025 s & 0.029 s  \\
            \addlinespace 
		IRAS & 18.12 s & 19.82 s & 20.69 s & 20.41 s & 22.97 s  \\
            \addlinespace 
		Optimal Solution & 4620.16 s & - & - & - & - \\
		\bottomrule
	\end{tabularx}
\end{table}




 
\subsection{Effect of Different Flow Period On Performance}
\begin{figure}[!htbp]
	\begin{minipage}{0.5\textwidth}
		\centering
		\includegraphics[width=0.7\linewidth]{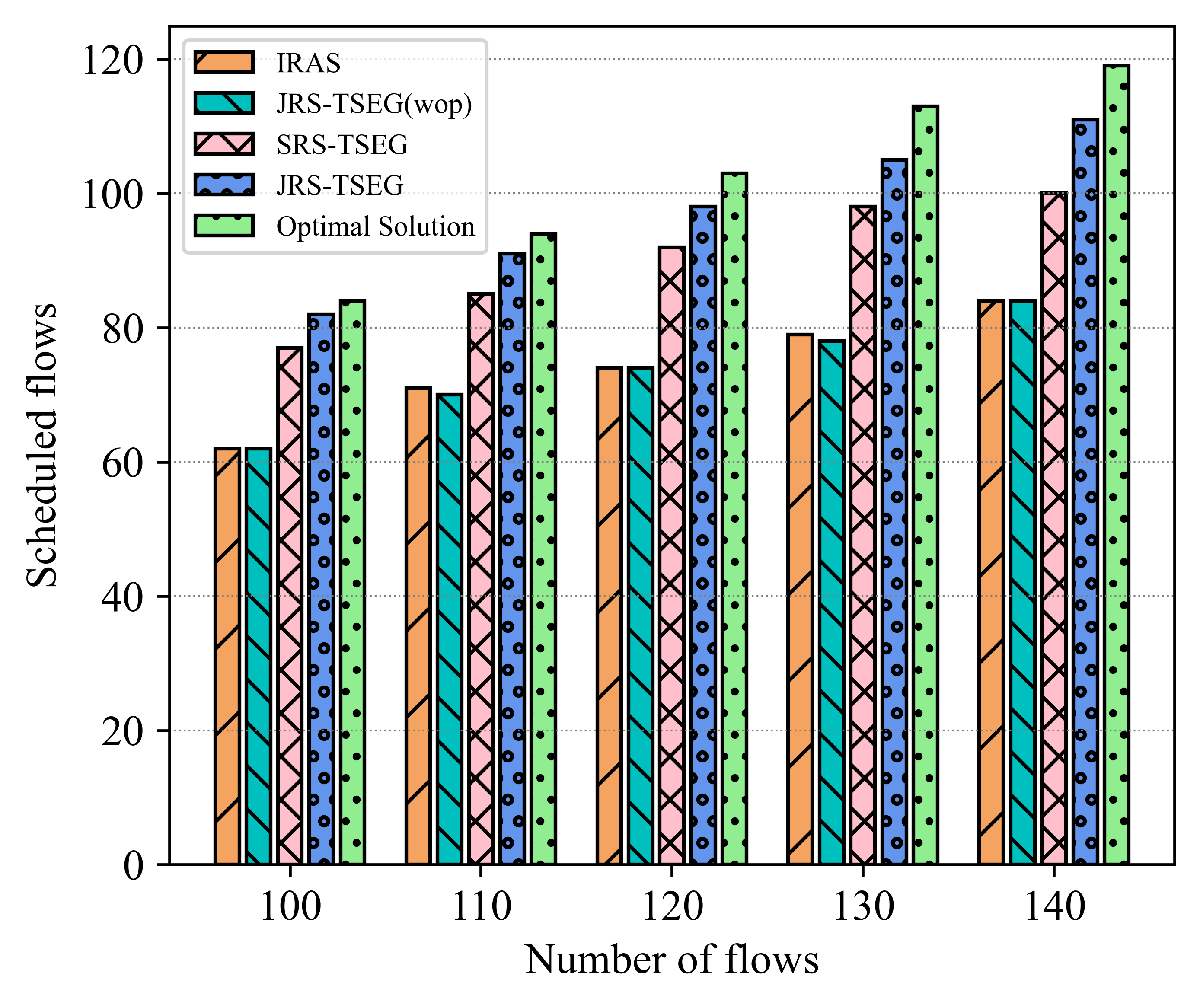}
		\caption{The comparison of scheduled flows among different algorithms in the ring topology with period set $\{60, 480\}$ and $\alpha=0.4$, $\delta=0.6$.}
		\label{fz4_1}
	\end{minipage}
\end{figure}

The period is an inherent attribute of flows and, at the same time, an important constraint in scheduling problems. In this section, we explore the effect of the flow period on scheduling performance. Here, we transform the original configurations, changing the period from $120 \mu s$ and $240 \mu s$ time slots to $60 \mu s$ and $480 \mu s$, respectively.
The parameters in Fig. \ref{fz4_1} are set to ring topology, period set $\{60, 480\}$ (unit: $\mu s$), and the flow proportions $\alpha=0.4$, $\delta=0.6$ respectively. 

Fig. \ref{fz4_1} illustrates the scheduling quality comparison of the different methods, where our method decreases to 90\% of the optimal solution compared to the case of the period set $\{60, 120, 240, 480\}$. 
Our method exhibits a performance gap of approximately 30\% compared to IRAS and JRS-TSEG (wop). However, when compared to SRS-TSEG, the performance gap narrows to 7\%. 
This is mainly attributed to a reduction in the number of small period flow types. Due to the large gap between period $60 \mu s$ and period $480 \mu s$, the weight values of each time slot on the link are not much different after updating, resulting in no significant difference in the weight values of the path between any two nodes, which is easily consistent with the path selected by SRS-TSEG.


Table \ref{runtime_4_2} shows the comparison of runtime among different algorithms.
Compared to Table \ref{ring_runtime}, the overall runtime of IRAS is larger and the run time of our method and SRS-TSEG is smaller because scheduling a flow with a period of $480 \mu s$ is more time-consuming than scheduling other flows.
In summary, the increase in the number of small period flows will promote the gap, but the decrease in the number of period types will change the potentially conflicting period types, thus affecting the gap between our method and SRS-TSEG.

\begin{table}
	\centering
	\caption{The average running time of different algorithms}
	\label{runtime_4_2}
	\begin{tabularx}{0.5\textwidth}{m{1.1cm} m{1.1cm} m{1.1cm} m{1.1cm} m{1.1cm} m{1.1cm}} 
		\toprule
		& 100 flows & 110 flows & 120 flows & 130 flows & 140 flows \\
		\midrule
		JRS-TSEG & 0.64 s & 0.75 s & 0.79 s & 0.89 s & 1.02 s  \\
            \addlinespace 
            SRS-TSEG & 3.09 s & 3.37 s & 4.66 s & 5.15 s & 5.61 s  \\
            \addlinespace 
            JRS-TSEG(wop) & 0.009 s & 0.022 s & 0.021 s & 0.015 s & 0.024 s  \\
            \addlinespace 
		IRAS & 5.78 s & 10.51 s & 12.84 s & 13.25 s & 13.61 s  \\
            \addlinespace 
		Optimal Solution & 2681.75 s & - & - & - & - \\
		\bottomrule
	\end{tabularx}
\end{table}

\section{Conclusion}\label{Conclusion}
In this paper, we study the problem of co-optimizing routing and scheduling in an online environment. We design a time-slot expanded graph to jointly represent the temporal and spatial resources in Time-Triggered Ethernet, transforming the routing and scheduling problem into a routing problem in the time-slot expanded graph. Based on the time-slot expanded graph, we model the routing and scheduling problem as an integer linear programming to provide a performance upper bound. We then propose a dynamic weighting method to add weight to each link in the graph, revealing the potential conflicts between flows. We further recognize that selecting the minimum-weight end-to-end path reduces the impact of the current flow's resource allocation on future flows. 
Thus, we design a routing algorithm that satisfies constraints for TTEthernet, aiming to find a path with the minimum weight, so that the potential conflicts between flows are eliminated. Our simulation results show that our approach runs $>400$ times faster than a standard ILP solver, while the gap of the successfully scheduled number of flows between our scheme and the optimal ILP solver is only 2\%. Besides, our scheme can improve the successfully scheduled number of flows to more than 18\% as compared to baselines. In our future work, we will study the problem of online routing and scheduling in deterministic networks.

\vfill

\end{document}